\documentclass[longauth]{aa}
\usepackage{amssymb,amsmath,amsfonts}
\usepackage{graphicx}
\usepackage{color}
\usepackage{ulem}             
\usepackage{multicol}
\usepackage{caption}
\usepackage{pgf,tikz}
\usetikzlibrary{arrows}
\usetikzlibrary{trees}
\usetikzlibrary{snakes}
\usetikzlibrary{shapes}
\usetikzlibrary{matrix}
\usetikzlibrary{calc}
\usetikzlibrary{positioning}
\usetikzlibrary{chains}
\usetikzlibrary{shadows}

\tikzstyle{Point} = [fill, radius=0.08]
\tikzstyle{BigPoint} = [fill, radius=0.13]
\tikzstyle{Leaf} = [color = gray]
\tikzstyle{Line1} = [dashed]
\tikzstyle{Line2} = [dotted, ultra thick]\usepackage{colortbl}

\usepackage{xcolor}

\newcommand\default{{\textit{default}} }
\newcommand\highm{{\textit{high-mass}} }
\newcommand\final{{\textit{final}} }

\newcommand{\be}{\begin{equation}}
\newcommand{\ee}{\end{equation}}

\usepackage[varg]{txfonts}
\usepackage{natbib}

\begin{document}

\title{Photo-dynamical characterisation of the TOI-178 resonant chain
}
\subtitle{ Exploring the robustness of TTV and RV mass characterisation}
\titlerunning{Photo-dynamical characterisation of the TOI-178 resonant chain}

\author{
A. Leleu\inst{1,2} $^{\href{https://orcid.org/0000-0003-2051-7974}{\protect\includegraphics[scale=0.5]{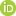}}}$, 
J.-B. Delisle\inst{1} $^{\href{https://orcid.org/0000-0001-5844-9888}{\protect\includegraphics[scale=0.5]{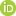}}}$, 
L. Delrez\inst{3,4,5} $^{\href{https://orcid.org/0000-0001-6108-4808}{\protect\includegraphics[scale=0.5]{figures/orcid.jpg}}}$, 
E. M. Bryant\inst{6} $^{\href{https://orcid.org/0000-0001-7904-4441}{\protect\includegraphics[scale=0.5]{figures/orcid.jpg}}}$, 
A. Brandeker\inst{7} $^{\href{https://orcid.org/0000-0002-7201-7536}{\protect\includegraphics[scale=0.5]{figures/orcid.jpg}}}$, 
H. P. Osborn\inst{8,9} $^{\href{https://orcid.org/0000-0002-4047-4724}{\protect\includegraphics[scale=0.5]{figures/orcid.jpg}}}$, 
N. Hara\inst{1} $^{\href{https://orcid.org/0000-0001-9232-3314}{\protect\includegraphics[scale=0.5]{figures/orcid.jpg}}}$, 
T. G. Wilson\inst{10} $^{\href{https://orcid.org/0000-0001-8749-1962}{\protect\includegraphics[scale=0.5]{figures/orcid.jpg}}}$, 
N. Billot\inst{1} $^{\href{https://orcid.org/0000-0003-3429-3836}{\protect\includegraphics[scale=0.5]{figures/orcid.jpg}}}$, 
M. Lendl\inst{1} $^{\href{https://orcid.org/0000-0001-9699-1459}{\protect\includegraphics[scale=0.5]{figures/orcid.jpg}}}$, 
D. Ehrenreich\inst{1,11} $^{\href{https://orcid.org/0000-0001-9704-5405}{\protect\includegraphics[scale=0.5]{figures/orcid.jpg}}}$, 
H. Chakraborty\inst{12}, 
M. N. Günther\inst{13} $^{\href{https://orcid.org/0000-0002-3164-9086}{\protect\includegraphics[scale=0.5]{figures/orcid.jpg}}}$, 
M. J. Hooton\inst{14} $^{\href{https://orcid.org/0000-0003-0030-332X}{\protect\includegraphics[scale=0.5]{figures/orcid.jpg}}}$, 
Y. Alibert\inst{8,2} $^{\href{https://orcid.org/0000-0002-4644-8818}{\protect\includegraphics[scale=0.5]{figures/orcid.jpg}}}$, 
R. Alonso\inst{15,16} $^{\href{https://orcid.org/0000-0001-8462-8126}{\protect\includegraphics[scale=0.5]{figures/orcid.jpg}}}$, 
D. R. Alves\inst{17} $^{\href{https://orcid.org/0000-0002-5619-2502}{\protect\includegraphics[scale=0.5]{figures/orcid.jpg}}}$, 
D. R. Anderson\inst{18} $^{\href{https://orcid.org/0000-0001-7416-7522}{\protect\includegraphics[scale=0.5]{figures/orcid.jpg}}}$, 
I. Apergis\inst{19} $^{\href{https://orcid.org/0009-0004-7473-4573}{\protect\includegraphics[scale=0.5]{figures/orcid.jpg}}}$, 
D. Armstrong\inst{19} $^{\href{https://orcid.org/0000-0002-5080-4117}{\protect\includegraphics[scale=0.5]{figures/orcid.jpg}}}$, 
T. Bárczy\inst{20} $^{\href{https://orcid.org/0000-0002-7822-4413}{\protect\includegraphics[scale=0.5]{figures/orcid.jpg}}}$, 
D. Barrado Navascues\inst{21} $^{\href{https://orcid.org/0000-0002-5971-9242}{\protect\includegraphics[scale=0.5]{figures/orcid.jpg}}}$, 
S. C. C. Barros\inst{22,23} $^{\href{https://orcid.org/0000-0003-2434-3625}{\protect\includegraphics[scale=0.5]{figures/orcid.jpg}}}$, 
M. P. Battley\inst{12} $^{\href{https://orcid.org/0000-0002-1357-9774}{\protect\includegraphics[scale=0.5]{figures/orcid.jpg}}}$, 
W. Baumjohann\inst{24} $^{\href{https://orcid.org/0000-0001-6271-0110}{\protect\includegraphics[scale=0.5]{figures/orcid.jpg}}}$, 
D. Bayliss\inst{19} $^{\href{https://orcid.org/0000-0001-6023-1335}{\protect\includegraphics[scale=0.5]{figures/orcid.jpg}}}$, 
T. Beck\inst{2}, 
W. Benz\inst{2,8} $^{\href{https://orcid.org/0000-0001-7896-6479}{\protect\includegraphics[scale=0.5]{figures/orcid.jpg}}}$, 
L. Borsato\inst{25} $^{\href{https://orcid.org/0000-0003-0066-9268}{\protect\includegraphics[scale=0.5]{figures/orcid.jpg}}}$, 
C. Broeg\inst{2,8} $^{\href{https://orcid.org/0000-0001-5132-2614}{\protect\includegraphics[scale=0.5]{figures/orcid.jpg}}}$, 
M. R. Burleigh\inst{26} $^{\href{https://orcid.org/0000-0003-0684-7803}{\protect\includegraphics[scale=0.5]{figures/orcid.jpg}}}$, 
S. L. Casewell\inst{26} $^{\href{https://orcid.org/0000-0003-2478-0120}{\protect\includegraphics[scale=0.5]{figures/orcid.jpg}}}$, 
A. Collier Cameron\inst{27} $^{\href{https://orcid.org/0000-0002-8863-7828}{\protect\includegraphics[scale=0.5]{figures/orcid.jpg}}}$, 
A. C. M. Correia\inst{28}, 
Sz. Csizmadia\inst{29} $^{\href{https://orcid.org/0000-0001-6803-9698}{\protect\includegraphics[scale=0.5]{figures/orcid.jpg}}}$, 
P. E. Cubillos\inst{30,24}, 
M. B. Davies\inst{31} $^{\href{https://orcid.org/0000-0001-6080-1190}{\protect\includegraphics[scale=0.5]{figures/orcid.jpg}}}$, 
M. Deleuil\inst{32} $^{\href{https://orcid.org/0000-0001-6036-0225}{\protect\includegraphics[scale=0.5]{figures/orcid.jpg}}}$, 
A. Deline\inst{1}, 
O. D. S. Demangeon\inst{22,23} $^{\href{https://orcid.org/0000-0001-7918-0355}{\protect\includegraphics[scale=0.5]{figures/orcid.jpg}}}$, 
B.-O. Demory\inst{8,2} $^{\href{https://orcid.org/0000-0002-9355-5165}{\protect\includegraphics[scale=0.5]{figures/orcid.jpg}}}$, 
A. Derekas\inst{33}, 
B. Edwards\inst{34}, 
A. Erikson\inst{29}, 
A. Fortier\inst{2,8} $^{\href{https://orcid.org/0000-0001-8450-3374}{\protect\includegraphics[scale=0.5]{figures/orcid.jpg}}}$, 
L. Fossati\inst{24} $^{\href{https://orcid.org/0000-0003-4426-9530}{\protect\includegraphics[scale=0.5]{figures/orcid.jpg}}}$, 
M. Fridlund\inst{35,36} $^{\href{https://orcid.org/0000-0002-0855-8426}{\protect\includegraphics[scale=0.5]{figures/orcid.jpg}}}$, 
D. Gandolfi\inst{37} $^{\href{https://orcid.org/0000-0001-8627-9628}{\protect\includegraphics[scale=0.5]{figures/orcid.jpg}}}$, 
K. Gazeas\inst{38}, 
E. Gillen\inst{39} $^{\href{https://orcid.org/0000-0003-2851-3070}{\protect\includegraphics[scale=0.5]{figures/orcid.jpg}}}$, 
M. Gillon\inst{3} $^{\href{https://orcid.org/0000-0003-1462-7739}{\protect\includegraphics[scale=0.5]{figures/orcid.jpg}}}$, 
M. R. Goad\inst{26}, 
M. Güdel\inst{40}, 
F. Hawthorn\inst{19} $^{\href{https://orcid.org/0000-0002-8675-182X}{\protect\includegraphics[scale=0.5]{figures/orcid.jpg}}}$, 
A. Heitzmann\inst{1} $^{\href{https://orcid.org/0000-0002-8091-7526}{\protect\includegraphics[scale=0.5]{figures/orcid.jpg}}}$, 
Ch. Helling\inst{24,41}, 
K. G. Isaak\inst{13} $^{\href{https://orcid.org/0000-0001-8585-1717}{\protect\includegraphics[scale=0.5]{figures/orcid.jpg}}}$, 
J. S. Jenkins\inst{42} $^{\href{https://orcid.org/0000-0003-2733-8725}{\protect\includegraphics[scale=0.5]{figures/orcid.jpg}}}$, 
J. M. Jenkins\inst{43}, 
A. Kendall\inst{26} $^{\href{https://orcid.org/0009-0006-0719-9229}{\protect\includegraphics[scale=0.5]{figures/orcid.jpg}}}$, 
L. L. Kiss\inst{44,45}, 
J. Korth\inst{46}, 
K. W. F. Lam\inst{29} $^{\href{https://orcid.org/0000-0002-9910-6088}{\protect\includegraphics[scale=0.5]{figures/orcid.jpg}}}$, 
J. Laskar\inst{47} $^{\href{https://orcid.org/0000-0003-2634-789X}{\protect\includegraphics[scale=0.5]{figures/orcid.jpg}}}$, 
D. W. Latham\inst{48}, 
A. Lecavelier des Etangs\inst{49} $^{\href{https://orcid.org/0000-0002-5637-5253}{\protect\includegraphics[scale=0.5]{figures/orcid.jpg}}}$, 
D. Magrin\inst{25} $^{\href{https://orcid.org/0000-0003-0312-313X}{\protect\includegraphics[scale=0.5]{figures/orcid.jpg}}}$, 
P. F. L. Maxted\inst{50} $^{\href{https://orcid.org/0000-0003-3794-1317}{\protect\includegraphics[scale=0.5]{figures/orcid.jpg}}}$, 
J. McCormac\inst{19} $^{\href{https://orcid.org/0000-0003-1631-4170}{\protect\includegraphics[scale=0.5]{figures/orcid.jpg}}}$, 
C. Mordasini\inst{2,8}, 
M. Moyano\inst{18} $^{\href{https://orcid.org/0000-0002-7927-9555}{\protect\includegraphics[scale=0.5]{figures/orcid.jpg}}}$, 
V. Nascimbeni\inst{25} $^{\href{https://orcid.org/0000-0001-9770-1214}{\protect\includegraphics[scale=0.5]{figures/orcid.jpg}}}$, 
G. Olofsson\inst{7} $^{\href{https://orcid.org/0000-0003-3747-7120}{\protect\includegraphics[scale=0.5]{figures/orcid.jpg}}}$, 
A. Osborn\inst{51} $^{\href{https://orcid.org/0000-0002-5899-7750}{\protect\includegraphics[scale=0.5]{figures/orcid.jpg}}}$, 
R. Ottensamer\inst{40}, 
I. Pagano\inst{52} $^{\href{https://orcid.org/0000-0001-9573-4928}{\protect\includegraphics[scale=0.5]{figures/orcid.jpg}}}$, 
E. Pallé\inst{15,16} $^{\href{https://orcid.org/0000-0003-0987-1593}{\protect\includegraphics[scale=0.5]{figures/orcid.jpg}}}$, 
G. Peter\inst{53} $^{\href{https://orcid.org/0000-0001-6101-2513}{\protect\includegraphics[scale=0.5]{figures/orcid.jpg}}}$, 
G. Piotto\inst{25,54} $^{\href{https://orcid.org/0000-0002-9937-6387}{\protect\includegraphics[scale=0.5]{figures/orcid.jpg}}}$, 
D. Pollacco\inst{10}, 
D. Queloz\inst{55,14} $^{\href{https://orcid.org/0000-0002-3012-0316}{\protect\includegraphics[scale=0.5]{figures/orcid.jpg}}}$, 
R. Ragazzoni\inst{25,54} $^{\href{https://orcid.org/0000-0002-7697-5555}{\protect\includegraphics[scale=0.5]{figures/orcid.jpg}}}$, 
N. Rando\inst{13}, 
H. Rauer\inst{29,56} $^{\href{https://orcid.org/0000-0002-6510-1828}{\protect\includegraphics[scale=0.5]{figures/orcid.jpg}}}$, 
I. Ribas\inst{57,58} $^{\href{https://orcid.org/0000-0002-6689-0312}{\protect\includegraphics[scale=0.5]{figures/orcid.jpg}}}$, 
G. Ricker\inst{9}, 
S. Saha\inst{42} $^{\href{https://orcid.org/0000-0001-8018-0264}{\protect\includegraphics[scale=0.5]{figures/orcid.jpg}}}$, 
N. C. Santos\inst{22,23} $^{\href{https://orcid.org/0000-0003-4422-2919}{\protect\includegraphics[scale=0.5]{figures/orcid.jpg}}}$, 
G. Scandariato\inst{52} $^{\href{https://orcid.org/0000-0003-2029-0626}{\protect\includegraphics[scale=0.5]{figures/orcid.jpg}}}$, 
S. Seager\inst{9,59,60} $^{\href{https://orcid.org/0000-0002-6892-6948}{\protect\includegraphics[scale=0.5]{figures/orcid.jpg}}}$, 
D. Ségransan\inst{1} $^{\href{https://orcid.org/0000-0003-2355-8034}{\protect\includegraphics[scale=0.5]{figures/orcid.jpg}}}$, 
A. E. Simon\inst{2,8} $^{\href{https://orcid.org/0000-0001-9773-2600}{\protect\includegraphics[scale=0.5]{figures/orcid.jpg}}}$, 
A. M. S. Smith\inst{29} $^{\href{https://orcid.org/0000-0002-2386-4341}{\protect\includegraphics[scale=0.5]{figures/orcid.jpg}}}$, 
S. G. Sousa\inst{22} $^{\href{https://orcid.org/0000-0001-9047-2965}{\protect\includegraphics[scale=0.5]{figures/orcid.jpg}}}$, 
M. Stalport\inst{4,3}, 
S. Sulis\inst{32} $^{\href{https://orcid.org/0000-0001-8783-526X}{\protect\includegraphics[scale=0.5]{figures/orcid.jpg}}}$, 
Gy. M. Szabó\inst{33,61} $^{\href{https://orcid.org/0000-0002-0606-7930}{\protect\includegraphics[scale=0.5]{figures/orcid.jpg}}}$, 
S. Udry\inst{1} $^{\href{https://orcid.org/0000-0001-7576-6236}{\protect\includegraphics[scale=0.5]{figures/orcid.jpg}}}$, 
V. Van Grootel\inst{4} $^{\href{https://orcid.org/0000-0003-2144-4316}{\protect\includegraphics[scale=0.5]{figures/orcid.jpg}}}$, 
R. Vanderspek\inst{9}, 
J. Venturini\inst{1} $^{\href{https://orcid.org/0000-0001-9527-2903}{\protect\includegraphics[scale=0.5]{figures/orcid.jpg}}}$, 
E. Villaver\inst{15,16}, 
J. I. Vinés\inst{18} $^{\href{https://orcid.org/0000-0002-1896-2377}{\protect\includegraphics[scale=0.5]{figures/orcid.jpg}}}$, 
N. A. Walton\inst{62} $^{\href{https://orcid.org/0000-0003-3983-8778}{\protect\includegraphics[scale=0.5]{figures/orcid.jpg}}}$, 
R. G. West\inst{19} $^{\href{https://orcid.org/0000-0001-6604-5533}{\protect\includegraphics[scale=0.5]{figures/orcid.jpg}}}$, 
J. Winn\inst{63} $^{\href{https://orcid.org/0000-0002-4265-047X}{\protect\includegraphics[scale=0.5]{figures/orcid.jpg}}}$, 
T. Zivave\inst{19}
}
\authorrunning{A. Leleu et al}

\institute{
\label{inst:1} Observatoire astronomique de l'Université de Genève, Chemin Pegasi 51, 1290 Versoix, Switzerland \and
\label{inst:2} Weltraumforschung und Planetologie, Physikalisches Institut, University of Bern, Gesellschaftsstrasse 6, 3012 Bern, Switzerland \and
\label{inst:3} Astrobiology Research Unit, Université de Liège, Allée du 6 Août 19C, B-4000 Liège, Belgium \and
\label{inst:4} Space sciences, Technologies and Astrophysics Research (STAR) Institute, Université de Liège, Allée du 6 Août 19C, 4000 Liège, Belgium \and
\label{inst:5} Institute of Astronomy, KU Leuven, Celestijnenlaan 200D, 3001 Leuven, Belgium \and
\label{inst:6} Mullard Space Science Laboratory, University College London, Holmbury St Mary, Dorking, Surrey, RH5 6NT, UK \and
\label{inst:7} Department of Astronomy, Stockholm University, AlbaNova University Center, 10691 Stockholm, Sweden \and
\label{inst:8} Center for Space and Habitability, University of Bern, Gesellschaftsstrasse 6, 3012 Bern, Switzerland \and
\label{inst:9} Department of Physics and Kavli Institute for Astrophysics and Space Research, Massachusetts Institute of Technology, Cambridge, MA 02139, USA \and
\label{inst:10} Department of Physics, University of Warwick, Gibbet Hill Road, Coventry CV4 7AL, United Kingdom \and
\label{inst:11} Centre Vie dans l’Univers, Faculté des sciences, Université de Genève, Quai Ernest-Ansermet 30, 1211 Genève 4, Switzerland \and
\label{inst:12} Observatoire Astronomique de l’Université de Genève, Chemin Pegasi 51, CH-1290 Versoix, Switzerland \and
\label{inst:13} European Space Agency (ESA), European Space Research and Technology Centre (ESTEC), Keplerlaan 1, 2201 AZ Noordwijk, The Netherlands \and
\label{inst:14} Cavendish Laboratory, JJ Thomson Avenue, Cambridge CB3 0HE, UK \and
\label{inst:15} Instituto de Astrofísica de Canarias, Vía Láctea s/n, 38200 La Laguna, Tenerife, Spain \and
\label{inst:16} Departamento de Astrofísica, Universidad de La Laguna, Astrofísico Francisco Sanchez s/n, 38206 La Laguna, Tenerife, Spain \and
\label{inst:17} Departamento de Astronom\'{i}a, Universidad de Chile, Casilla 36-D, Santiago, Chile \and
\label{inst:18} Instituto de Astronomía, Universidad Católica del Norte, Angamos 0610, Antofagasta 1270709, Chile \and
\label{inst:19} Department of Physics, University of Warwick, Gibbet Hill Road, Coventry CV4 7AL, UK \and
\label{inst:20} Admatis, 5. Kandó Kálmán Street, 3534 Miskolc, Hungary \and
\label{inst:21} Depto. de Astrofísica, Centro de Astrobiología (CSIC-INTA), ESAC campus, 28692 Villanueva de la Cañada (Madrid), Spain \and
\label{inst:22} Instituto de Astrofisica e Ciencias do Espaco, Universidade do Porto, CAUP, Rua das Estrelas, 4150-762 Porto, Portugal \and
\label{inst:23} Departamento de Fisica e Astronomia, Faculdade de Ciencias, Universidade do Porto, Rua do Campo Alegre, 4169-007 Porto, Portugal \and
\label{inst:24} Space Research Institute, Austrian Academy of Sciences, Schmiedlstrasse 6, A-8042 Graz, Austria \and
\label{inst:25} INAF, Osservatorio Astronomico di Padova, Vicolo dell'Osservatorio 5, 35122 Padova, Italy \and
\label{inst:26} Centre for Exoplanet Research, School of Physics and Astronomy, University of Leicester, Leicester LE1 7RH, UK \and
\label{inst:27} Centre for Exoplanet Science, SUPA School of Physics and Astronomy, University of St Andrews, North Haugh, St Andrews KY16 9SS, UK \and
\label{inst:28} CFisUC, Department of Physics, University of Coimbra, 3004-516 Coimbra, Portugal \and
\label{inst:29} Institute of Planetary Research, German Aerospace Center (DLR), Rutherfordstrasse 2, 12489 Berlin, Germany \and
\label{inst:30} INAF, Osservatorio Astrofisico di Torino, Via Osservatorio, 20, I-10025 Pino Torinese To, Italy \and
\label{inst:31} Centre for Mathematical Sciences, Lund University, Box 118, 221 00 Lund, Sweden \and
\label{inst:32} Aix Marseille Univ, CNRS, CNES, LAM, 38 rue Frédéric Joliot-Curie, 13388 Marseille, France \and
\label{inst:33} ELTE Gothard Astrophysical Observatory, 9700 Szombathely, Szent Imre h. u. 112, Hungary \and
\label{inst:34} SRON Netherlands Institute for Space Research, Niels Bohrweg 4, 2333 CA Leiden, Netherlands \and
\label{inst:35} Leiden Observatory, University of Leiden, PO Box 9513, 2300 RA Leiden, The Netherlands \and
\label{inst:36} Department of Space, Earth and Environment, Chalmers University of Technology, Onsala Space Observatory, 439 92 Onsala, Sweden \and
\label{inst:37} Dipartimento di Fisica, Università degli Studi di Torino, via Pietro Giuria 1, I-10125, Torino, Italy \and
\label{inst:38} National and Kapodistrian University of Athens, Department of Physics, University Campus, Zografos GR-157 84, Athens, Greece \and
\label{inst:39} Astronomy Unit, Queen Mary University of London, Mile End Road, London E1 4NS, UK \and
\label{inst:40} Department of Astrophysics, University of Vienna, Türkenschanzstrasse 17, 1180 Vienna, Austria \and
\label{inst:41} Institute for Theoretical Physics and Computational Physics, Graz University of Technology, Petersgasse 16, 8010 Graz, Austria \and
\label{inst:42} Instituto de Estudios Astrofísicos, Facultad de Ingeniería y Ciencias, Universidad Diego Portales, Av. Ejército Libertador 441, Santiago, Chile \and
\label{inst:43} NASA Ames Research Center, Moffett Field, CA 94035, USA \and
\label{inst:44} Konkoly Observatory, Research Centre for Astronomy and Earth Sciences, 1121 Budapest, Konkoly Thege Miklós út 15-17, Hungary \and
\label{inst:45} ELTE E\"otv\"os Lor\'and University, Institute of Physics, P\'azm\'any P\'eter s\'et\'any 1/A, 1117 Budapest, Hungary \and
\label{inst:46} Lund Observatory, Division of Astrophysics, Department of Physics, Lund University, Box 43, 22100 Lund, Sweden \and
\label{inst:47} IMCCE, UMR8028 CNRS, Observatoire de Paris, PSL Univ., Sorbonne Univ., 77 av. Denfert-Rochereau, 75014 Paris, France \and
\label{inst:48} Center for Astrophysics, Harvard and Smithsonian, 60 Garden Street, Cambridge, MA 02138, USA \and
\label{inst:49} Institut d'astrophysique de Paris, UMR7095 CNRS, Université Pierre \& Marie Curie, 98bis blvd. Arago, 75014 Paris, France \and
\label{inst:50} Astrophysics Group, Lennard Jones Building, Keele University, Staffordshire, ST5 5BG, United Kingdom \and
\label{inst:51} Department of Physics and Astronomy, McMaster University, 1280 Main Street West, Hamilton, Ontario, L8S 4L8 \and
\label{inst:52} INAF, Osservatorio Astrofisico di Catania, Via S. Sofia 78, 95123 Catania, Italy \and
\label{inst:53} Institute of Optical Sensor Systems, German Aerospace Center (DLR), Rutherfordstrasse 2, 12489 Berlin, Germany \and
\label{inst:54} Dipartimento di Fisica e Astronomia "Galileo Galilei", Università degli Studi di Padova, Vicolo dell'Osservatorio 3, 35122 Padova, Italy \and
\label{inst:55} ETH Zurich, Department of Physics, Wolfgang-Pauli-Strasse 2, CH-8093 Zurich, Switzerland \and
\label{inst:56} Institut fuer Geologische Wissenschaften, Freie Universitaet Berlin, Maltheserstrasse 74-100,12249 Berlin, Germany \and
\label{inst:57} Institut de Ciencies de l'Espai (ICE, CSIC), Campus UAB, Can Magrans s/n, 08193 Bellaterra, Spain \and
\label{inst:58} Institut d'Estudis Espacials de Catalunya (IEEC), 08860 Castelldefels (Barcelona), Spain \and
\label{inst:59} Department of Earth, Atmospheric and Planetary Sciences, Massachusetts Institute of Technology, Cambridge, MA 02139, USA \and
\label{inst:60} Department of Aeronautics and Astronautics, MIT, 77 Massachusetts Avenue, Cambridge, MA 02139, USA \and
\label{inst:61} HUN-REN-ELTE Exoplanet Research Group, Szent Imre h. u. 112., Szombathely, H-9700, Hungary \and
\label{inst:62} Institute of Astronomy, University of Cambridge, Madingley Road, Cambridge, CB3 0HA, United Kingdom \and
\label{inst:63} Department of Astrophysical Sciences, Princeton University, Princeton, NJ 08544, USA
}

 \abstract
   {The TOI-178 system consists of a nearby late K-dwarf transited by six planets in the super-Earth to mini-Neptune regime, with radii ranging from $\sim$1.2 to 2.9 $R_{\oplus}$ and orbital periods between 1.9 and 20.7 days. All planets but the innermost one form a chain of Laplace resonances. 
   The fine-tuning and fragility of such orbital configurations ensure that no significant scattering or collision event has taken place since the {formation and migration} of the planets in the protoplanetary disc, hence providing important anchors for planet formation models.
   }
   {We aim to improve the characterisation of the architecture of this key system, and in particular the masses and radii of its planets. In addition, since this system is one of the few resonant chains that can be characterised by both photometry and radial velocities, we aim to use it as a test bench for the robustness of the planetary mass determination with each technique.   }
   {We perform a global analysis of all available photometry and radial velocity using photo-dynamical modelling of the light curve. We also try different sets of priors on the masses and eccentricity, as well as different stellar activity models, to study their effects on the masses estimated by transit timing variations and radial velocities.}
   {
   We show how stellar activity is preventing us from obtaining {a} robust mass estimation for the three outer planets using radial velocity data alone. 
   We also show that our joint photo-dynamical and radial velocity analysis resulted in {a} robust mass determination for planets $c$ to $g$, with precision of $\sim 12\%$ for the mass of planet $c$, and better than $10\%$ for planets $d$ to $g$. The new precisions on the radii range from $2$ to $3\%$. The understanding of this synergy between photometric and radial velocity measurements will be valuable during the PLATO mission. We also show that TOI-178 is indeed currently locked in the resonant configuration, librating around an equilibrium of the chain.
   }
  {}


\maketitle

\section{Introduction}

The observed architecture of planetary systems, defined as the orbit and composition of their planets, is the outcome of their formation in their proto-planetary disc and long-term evolution (typically Gyrs) after its dispersal. 
In this context, planetary systems observed in chains of Laplace resonances, where each consecutive pair of planets are in (or close to) a 2-body mean-motion resonance (MMR), 
are of particular interest. Indeed, the fine-tuning and fragility of such orbital configurations ensure that no significant scattering or collision event has taken place since the {end of the migration} of the planets in the protoplanetary disc \citep[e.g.][]{2016Natur.533..509M,Izidoro2017}. Hence, these systems are especially valuable for constraining the outcome of protoplanetary discs and provide important anchors for planet formation models. 

To date, chains of Laplace resonances have only been observed for a few systems: GJ 876 \citep{2010ApJ...719..890R}, Kepler-60 \citep{2016MNRAS.455L.104G}, Kepler-80 \citep{2016AJ....152..105M}, Kepler-223 \citep{2016Natur.533..509M}, TRAPPIST-1 \citep{2017Natur.542..456G,2017NatAs...1E.129L}, K2-138 \citep{Lopez2019}, TOI-178 (\citealt{Leleu2021}, hereafter L21), TOI-1136 \citep{Dai2023} and HD 110067 \citep{Luque2023}. All these systems, except GJ 876, are transiting, which provides an opportunity to observe the effect of planet-planet gravitational interactions, and thus constrain the masses and eccentricities of the planets via their transit timing variations (TTVs).
For stars that are bright enough, it is also possible to obtain radial velocity (RV) measurements, that can provide complementary constraints on the planetary masses and orbital parameters. Out of the transiting systems cited above, only K2-138, TOI-178, TOI-1136 and HD110067 have published RV measurements so far, the other ones being too faint in the visible ($V$-mag$\gtrsim$14). Having the possibility to measure planetary masses independently from RVs and TTVs is especially valuable to understand how choices of noise models and degeneracies between parameters affect the robustness of mass measurements for each technique.

Using data from \textit{Transiting Exoplanet Survey Satellite} (TESS, \citealt{2015JATIS...1a4003R}), 
\textit{CHaracterising ExOPlanets Satellite} (CHEOPS, \citealt{2021ExA....51..109B}) and \textit{Next Generation Transit Survey} \citep[NGTS,][]{wheatley2018ngts}, L21 showed that the nearby ($\sim$63 pc) late K-type star TOI-178 is a compact system of at least six transiting planets in the super-Earth to mini-Neptune regime, with radii ranging from $\sim$1.1 to 2.9 $R_{\oplus}$ and orbital periods of 1.91, 3.24, 6.56, 9.96, 15.23, and 20.71 days. The planetary radii were later refined by \cite{Delrez2023}, thereafter D23. The five outer planets form a 2:4:6:9:12 chain of Laplace resonances, while the innermost planets $b$ and $c$ are just wide of the 3:5 MMR, which could indicate that it was previously part of the chain but was then pulled away, possibly by tidal forces (L21).


Using RV measurements obtained with the Echelle SPectrograph for Rocky Exoplanets and Stable Spectroscopic Observations (ESPRESSO, \citealt{2021A&A...645A..96P}) installed at ESO's Very Large Telescope, L21 were also able to derive preliminary estimates for the masses of the planets, and thus their bulk densities (when combined with the radii inferred from the transit photometry). The planetary densities that they found show important variations from planet to planet, jumping for example from $\sim$1 to 0.2 $\rho_{\oplus}$ between planets $c$ and $d$. By doing a Bayesian internal structure analysis, they showed that the two innermost planets are likely to be mostly rocky, which could indicate that they have lost their primordial gas envelope through atmospheric escape, while all the other planets appear to contain significant amounts of water and/or gas (see also the independent internal structure analysis by \citealt{2022A&A...660A.102A}).
%
However, the planetary densities on which these results were based were constrained by a relatively low number of RV points (46 ESPRESSO points).
%

In this paper, we re-analyse the data presented in L21 and D23, to which we add CHEOPS observations taken in 2021, 2022 and 2023, TESS sector 69 (2023), and NGTS observations taken in 2021, see section \ref{sec:obs}. The whole analysis is performed by a photo-dynamical fit of all available photometry joint with the fit of available RV measurments, see section \ref{sec:data_ana}. We then present the result of our analysis in section \ref{sec:results}. In particular, we discuss the robustness of the extracted masses, exploring the effect of the mass-eccentricity degeneracy in TTVs and the effect of activity modelling in RV. Finally, we conclude in section \ref{sec:conclusion}.

\section{Data}
\label{sec:obs}

\begin{figure*}[!ht]
\begin{center}
\includegraphics[width=0.99\linewidth]{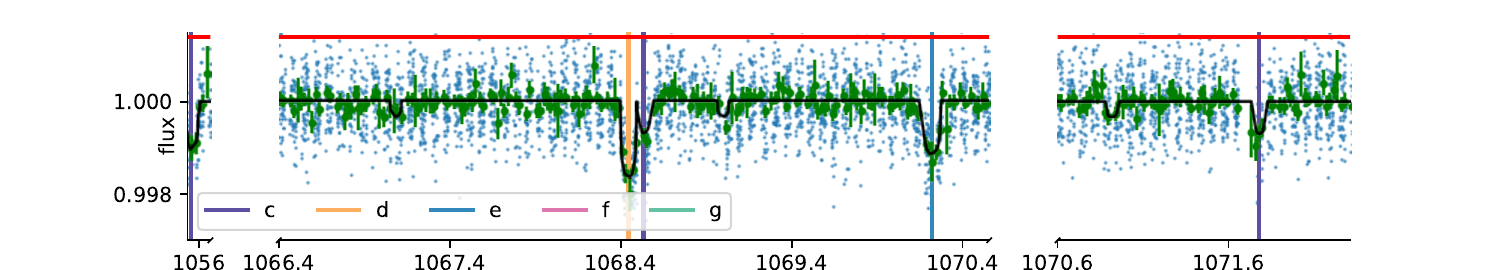}\\
\includegraphics[width=0.99\linewidth]{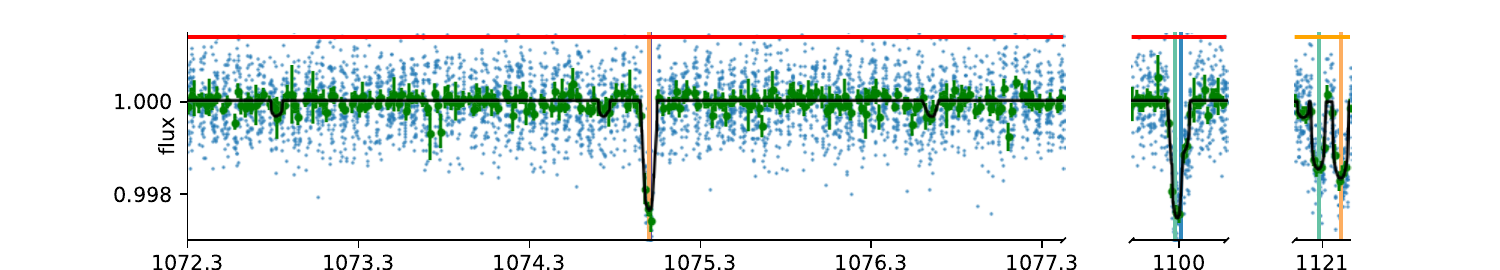}\\
\includegraphics[width=0.99\linewidth]{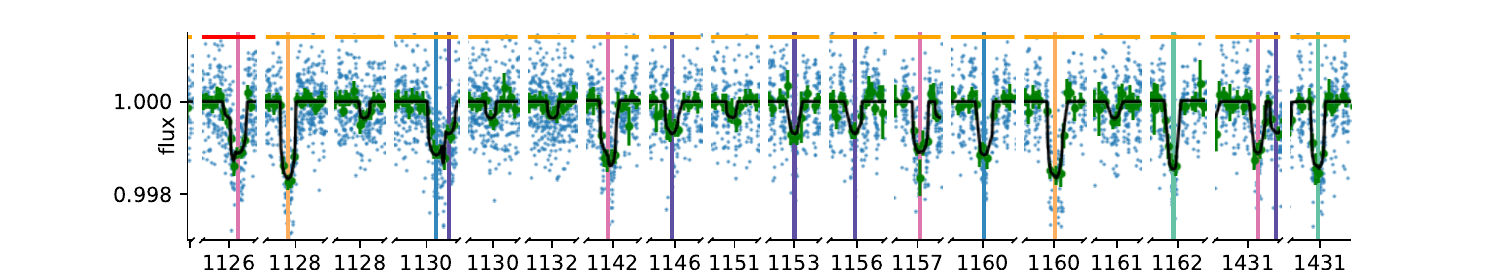}\\
\includegraphics[width=0.99\linewidth]{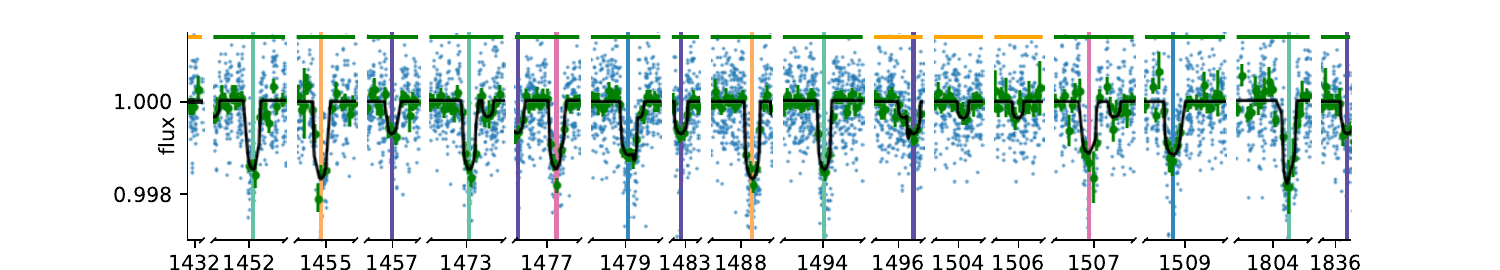}\\
\includegraphics[width=0.99\linewidth]{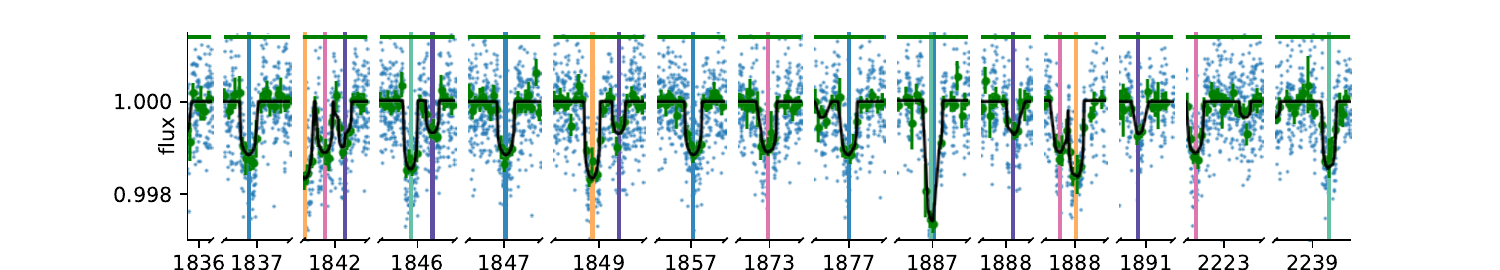}\\
\includegraphics[width=0.99\linewidth]{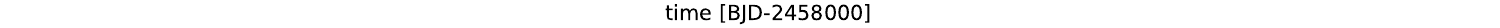}\\
\caption{\label{fig:CHEOPS_data} Detrended light curves from CHEOPS as described in sec \ref{sec:dataCHEOPS} and \ref{sec:data_modelling} Un-binned data are shown as blue points, and data in 30 min bins are shown as green circles. The best fitting transit model for the system is shown in black; the associated parameter values are from the final posterior shown in Tables \ref{tab:planet1} and \ref{tab:planet2}. Vertical lines indicate the planet that is transiting, unflagged transits are caused by planet $b$. Each line contains about six days of observation. Data published in L21 are indicated by a red upper line, the one published in D23 by an orange upper line, and new data by a green upper line. Raw data available online.
}
\end{center}
\end{figure*}

\begin{figure*}[!ht]
\begin{center}
\includegraphics[width=0.99\linewidth]{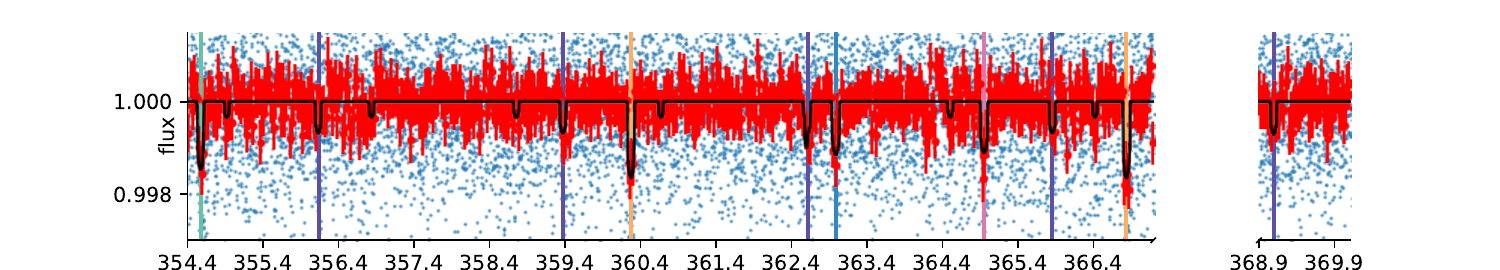}\\
\includegraphics[width=0.99\linewidth]{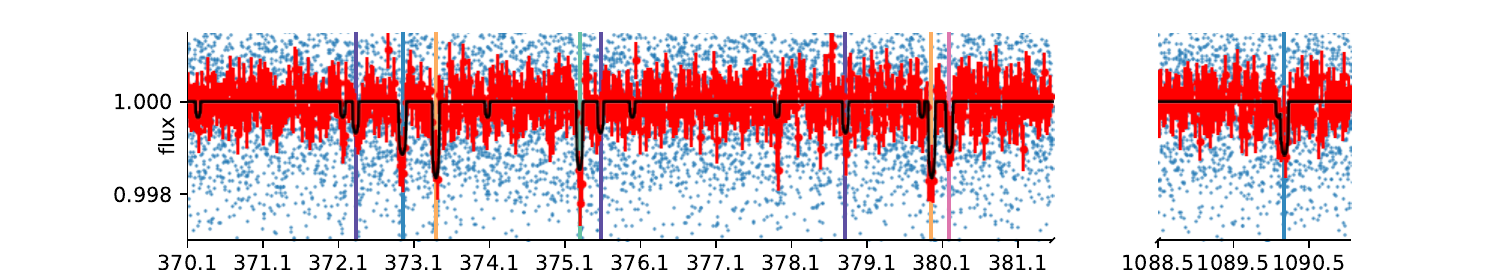}\\
\includegraphics[width=0.99\linewidth]{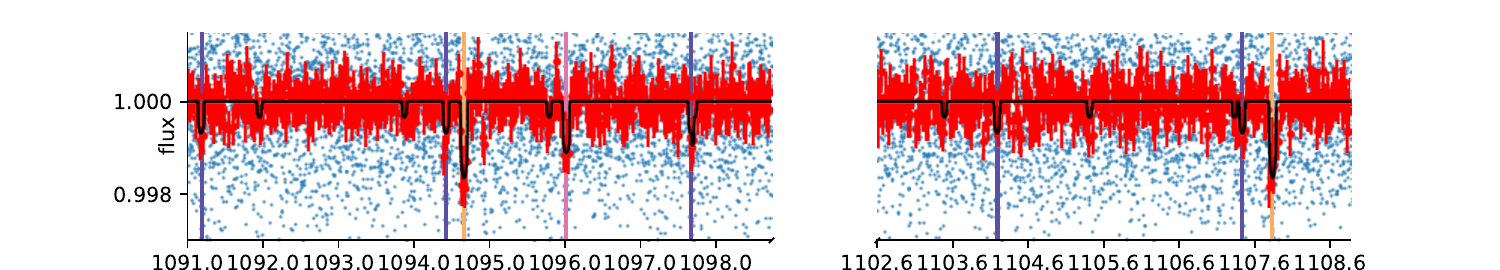}\\
\includegraphics[width=0.99\linewidth]{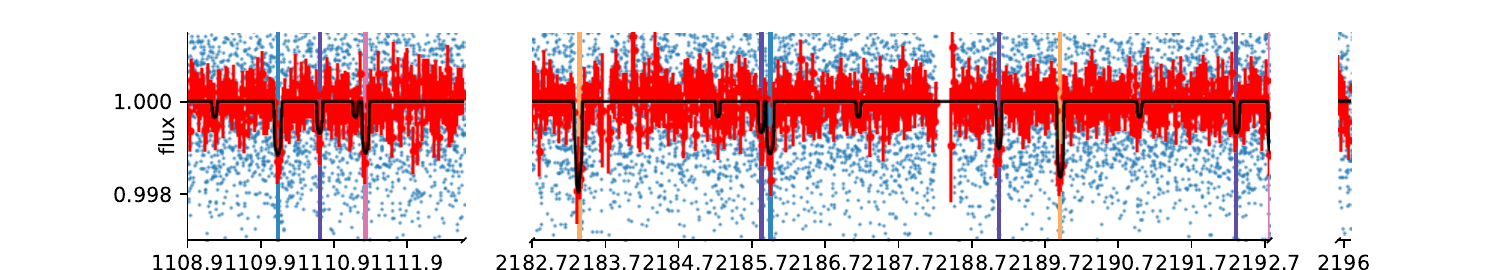}\\
\includegraphics[width=0.99\linewidth]{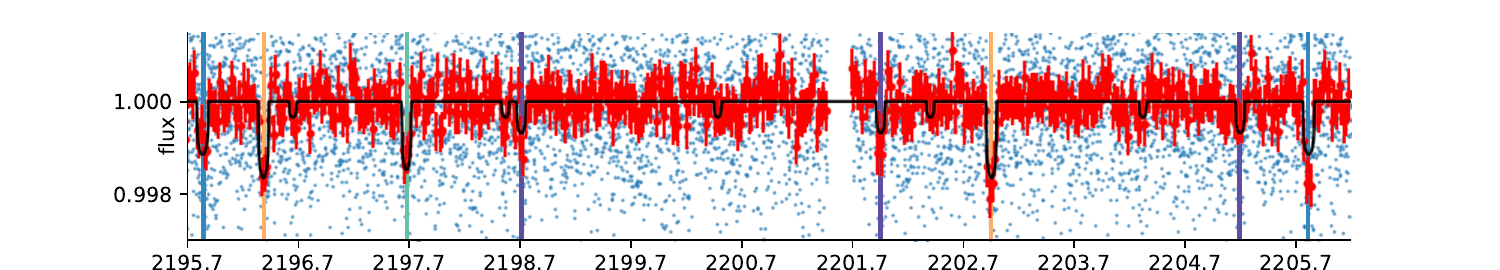}\\
\includegraphics[width=0.99\linewidth]{xlabel_Large.pdf}\\
\caption{\label{fig:TESS_data}  Detrended light curves from TESS as described in sec \ref{sec:dataTESS} and \ref{sec:data_modelling} Un-binned data are shown as blue points, and data in 30 min bins are shown as red circles. The best fitting transit model for the system is shown in black; the associated parameter values are from the final posterior shown in Tables \ref{tab:planet1} and \ref{tab:planet2}. Each line contains about 13 days of observation. 
Raw data available online.
}
\end{center}
\end{figure*}

\begin{figure*}[!ht]
\begin{center}
\includegraphics[width=0.65\linewidth]{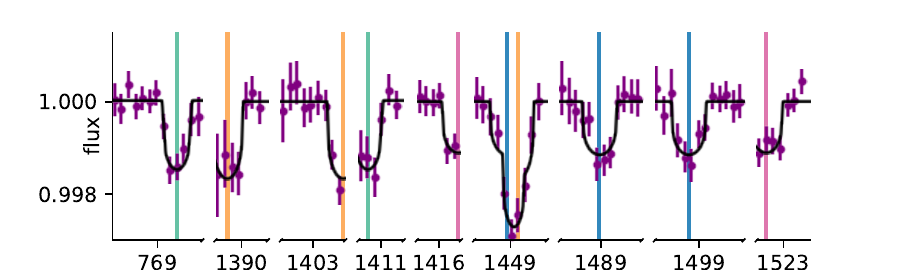}\\
\includegraphics[width=0.65\linewidth]{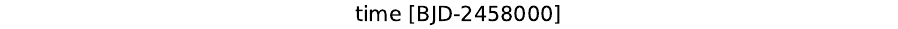}\\
\caption{\label{fig:NGTS_data}  Detrended light curves from NGTS as described in sec \ref{sec:dataNGTS} and \ref{sec:data_modelling}. Data in 30 min bins are shown as purple circles. The best fitting transit model for the system is shown in black; the associated parameter values are from the final posterior shown in Tables \ref{tab:planet1} and \ref{tab:planet2}. The line contains about 2 days of observation, each visit containing the data taken by respectively 7, 4, 3, 4, 4, 4, 3, 4 and 5 of the NGTS telescopes. Un-binned data are not shown because the scatter  is much larger than the than the flux range shown here. Raw data available online. 
}
\end{center}
\end{figure*}

In this study, we used the TESS, CHEOPS, ESPRESSO and NGTS data presented in L21 and D23. In addition, we use 27 new CHEOPS visits, 8 new NGTS observations and 4 EulerCam observations that were taken in order to monitor the TTVs of the 5 outer planets of TOI-178. The new data is available online (link). We also added the new data from the TESS sector 69.

\subsection{CHEOPS}
\label{sec:dataCHEOPS}

Following D23, the raw data of each visit were reduced with PIPE\footnote{\url{https://github.com/alphapsa/PIPE}} \citep{2022A&A...659L...4B}, a PSF photometry package developed specifically for CHEOPS. PIPE first uses a principal component analysis (PCA) approach to derive a PSF template library from the data series. The first five principal components (PCs) together with a constant background are then used to fit the individual PSFs of each image using a least-squares minimisation and measure the target's flux. The number of PCs to use is a trade-off between following systematic PSF changes and overfitting the noise. For faint stars such as TOI-178, the mean PSF (first PC) is sufficient for a good extraction, and attempts to model the PSF better with more PCs usually introduce noise in the extracted light curve. Some advantages of using PSF photometry rather than aperture photometry for faint targets are that: (1) the contributions to the signal of each pixel over the PSF are weighted according to noise so that higher S/N photometry can be extracted; (2) cosmic rays and bad pixels (both hot and telegraphic) are easier to filter out or give lower weight in the fitting process; (3) PSF photometry is less sensitive to contamination from nearby background stars; (4) the background is fit simultaneously with the PSF for the same pixels, which can be an advantage if there is some spatial structure. For each visit, we then filter out all the data for which the background contamination {rises} above 300 [electrons/pixel/exposure]. The detrended data is shown in Fig. \ref{fig:CHEOPS_data}.

\subsection{TESS}
\label{sec:dataTESS}

TESS \citep{2015JATIS...1a4003R} observed TOI-178 for the first time during Cycle 1/Sector 2 of its primary mission (22 August -- 20 September 2018). These data, obtained with a two-minute cadence, were previously presented in L21 and we include them in our global analysis. TESS observed again TOI-178 during Cycle 3/Sector 29 and  Cycle 5/Sector 69 of its extended mission, from 26 August to 22 September 2020 (presented in D23) and from 25 August to 20 September 2023. The data were processed with the TESS Science Processing Operations Center (SPOC) pipeline \citep{2016SPIE.9913E..3EJ} at NASA Ames Research Center. We retrieved the 2-minute cadence Pre-search Data Conditioning Simple Aperture Photometry (PDCSAP, \citealt{2012PASP..124..985S,2012PASP..124.1000S,2014PASP..126..100S}) from the Mikulski Archive for Space Telescopes\footnote{\url{https://archive.stsci.edu}} (MAST), using the default quality bitmask. The detrended data is shown in Fig. \ref{fig:TESS_data}.

\subsection{NGTS}
\label{sec:dataNGTS}

We also included in our global analysis the light curves obtained with the Next Generation Transit Survey (NGTS, \citealt{wheatley2018ngts}) that were previously published in L21, as well as eight additional transit observations that were obtained between the dates of 23 June 2021 and 3 November 2021. The NGTS photometric facility consists of twelve independently operated robotic telescopes each with a 20\,cm diameter aperture and a field-of-view of 8\,square-degrees. Using multiple NGTS telescopes to simultaneously observe the same star has been shown to yield vast improvements in the photometric precision compared to observing with a single telescope \citep{smith2020multicam, bryant2020multicam} and TOI-178 was observed using this high-precision multi-telescope observing mode. 
We observed three transits of TOI-178~d, two transits of TOI-178~e, two transits of TOI-178~f, and one transit of TOI-178~g. The NGTS observations were all performed using the custom NGTS filter (520 -- 890\,nm) at a cadence of 13\,seconds and were reduced using a custom aperture photometry pipeline which utilises the SEP package \citep{bertin96sextractor, Barbary2016} to perform the source extraction and also automatically selects comparison stars which are similar to TOI-178 in magnitude, colour, and CCD position using \textit{Gaia} DR2 \citep{GaiaCollaboration2018}. For more details on the reduction we refer the reader to \citet{bryant2020multicam}. The NGTS light curves used are displayed in Figure~\ref{fig:NGTS_data}.
We refer the reader to L21 and references therein for more information about the previously published NGTS data and their reduction.



\subsection{EulerCam}
\label{sec:EulerCam}

Multiple transits of planets d, e and f were observed with the EulerCam instrument between 2021-08-11 and 2021-10-20. EulerCam is a CCD imager installed at the Cassegrain focus of the 1.2m Leonhard Euler telescope at La Silla observatory \citep{lendl2012}. We used the broadband NGTS filter for our observations to maximise SNR. The data were reduced using the standard EulerCam pipeline, which performs basic image calibration and relative aperture photometry. The optimal aperture and reference stars are
selected by minimising the residual RMS of the final light curve. The EulerCam observations took place simultaneously to CHEOPS and - as expected for ground-based data -  show much lower precision (RMS in the 900-1800 ppm range). We eventually excluded them from our final analysis, but show them in Fig. \ref{fig:EulerCam} for completeness.

\subsection{ESPRESSO}
The RV data, presented in L21, consist of 46 ESPRESSO points. Each measurement was taken in high resolution (HR) mode with an integration time of 20 min using a single telescope (UT) and slow read-out (HR 21). The source on fibre B is the Fabry-Perot interferometer. Observations were made with a maximum airmass of 1.8 and a minimum 30$^\circ$ separation from the Moon. 

\section{Data analysis}
\label{sec:data_ana}

\subsection{Stellar properties}
\label{sec:star}

\renewcommand{\arraystretch}{1.33}

\begin{table}[hbt!]
\caption{Properties of the host star TOI-178.}
\begin{tabular}{l c c}
\hline 
\hline 
\textbf{Property (unit)} & \textbf{Value} & \textbf{Source} \\
\hline 
\multicolumn{2}{l}{\textit{Astrometric properties}} & \\
RA [J2000] & 00:29:12.49 & [1] \\
Dec [J2000] & $-$30:27:14.86 & [1] \\
$\mathrm{\mu_{RA}}$ [mas $\mathrm{yr}^{-1}$] & $150.032 \pm 0.028$ & [1] \\
$\mathrm{\mu_{Dec}}$ [mas $\mathrm{yr}^{-1}$] & $-87.132 \pm 0.030$ & [1] \\
Parallax [mas] & $15.900 \pm 0.031$ & [1] \\
Distance [pc] & $62.89 \pm 0.12$ & from parallax \\
\hline 
\multicolumn{2}{l}{\textit{Photometric magnitudes}} & \\
$G$ [mag] & $11.1575 \pm 0.0028$ & [1]\\
$G_{\rm BP}$ [mag] & $11.8398 \pm 0.0029$ & [1]\\
$G_{\rm RP}$ [mag] & $10.3602 \pm 0.0038$ & [1]\\
$J$ [mag] & $9.372 \pm 0.021$ & [2] \\
$H$ [mag] & $8.761 \pm 0.023$ & [2] \\
$K$ [mag] & $8.656 \pm 0.021$ & [2] \\
$W$1 [mag] & $8.573 \pm 0.022$ & [3]\\
$W$2 [mag] & $8.64 \pm 0.02$ & [3]\\
\hline 
\multicolumn{2}{l}{\textit{Spectroscopic and derived properties}} & \\
$T_{\mathrm{eff}}$ (K) & $4316 \pm 70$ & Spectroscopy [4] \\ 
log $g_\star$ (cgs) & $4.45 \pm 0.15$ & Spectroscopy [4] \\ 
$\mathrm{[Fe/H]}$ (dex) & $-0.23 \pm 0.05$ & Spectroscopy [4] \\
$v$ sin $i_\star$ (km $\mathrm{s}^{-1}$) & $1.5 \pm 0.3$ & Spectroscopy [4] \\
$R_\star$ ($R_\odot$) & $0.662 \pm 0.010$ & IRFM [5] \\
$M_\star$ ($M_\odot$) & $0.647_{-0.029}^{+0.030}$ & Isochrones [5] \\ 
$t_\star$ (Gyr) & $6.0_{-5.0}^{+6.8}$ & Isochrones [5] \\
$L_\star$ ($L_\odot$) & $0.136\pm0.010$ & from $R_\star$ and $T_{\mathrm{eff}}$ [5] \\
$\rho_\star$ ($\rho_\odot$) & $2.23\pm0.14$ & from $R_\star$ and $M_\star$ [5] \\ 
\hline 
\hline 
\end{tabular}
\textbf{References.} [1] \textit{Gaia} EDR3 \citep{GaiaCollaboration2021}; [2] 2MASS \citep{Skrutskie2006}; [3] WISe \citep{Wright2010}; [4] \cite{Leleu2021}; [5] \cite{Delrez2023}.
\label{tab:star}
\end{table}

In this study, we use the stellar properties that were updated by D23. These properties are summarized in Table \ref{tab:star}.

\subsection{Approach}
\label{sec:approach}

The transit timing variations expected for TOI-178 were explained in section 6.4 and illustrated in Fig. 14 of L21. A dominant feature of these TTVs is the effect of the proximity of 2-body MMRs \citep{Lithwick2012} which induce sinusoidal TTVs with a period of $\approx 260$ days (called super-period) on the 5 planets that are part of the resonant chain (planets $c$ to $g$).
Transit timing variations due to the proximity of 2-body MMRs are known to present a mass-eccentricity degeneracy at first order in the eccentricities \citep{Boue2012,Lithwick2012}. This degeneracy can be broken if either higher harmonics of the super-period \citep{HaLi2016}, the short-term chopping effect \citep{Deck2015}, or the long-term evolution in the resonance, can be constrained. 

To check the robustness of the masses we derive, we follow \cite{HaLi2017} and try out priors that pull the solution toward opposite directions of the degeneracy. We use their \textit{default} prior which is log-uniform in planet masses and uniform in eccentricities, and their  \textit{high-mass} prior which is uniform in planet masses and log-uniform in eccentricities. In addition, following \cite{Leleu2023}, we performed a third fit, using log-uniform mass prior and the \cite{Kipping2013} prior for the eccentricity: a $\beta$-distribution of parameters $\alpha=0.697$ and $\beta = 3.27$. The posterior associated with this set of priors is referred to as the  \textit{final} posterior. Then, following \cite{Leleu2023}, we quantify the robustness of the mass determination by studying the difference between the mass posteriors using the quantity $\Delta_{M}$:
\begin{equation}
\Delta_{M}=\max(\Delta_{M+},\Delta_{M-}),
\label{eq:deltaM}
\end{equation}
which estimates `at how many sigmas' of the final posterior the median of the default and highmass posteriors are. More precisely
\begin{equation}
\Delta_{M+} = \frac{ m_{high,0}-m_{final,0}}{m_{final,+\sigma}-m_{final,0}},
\label{eq:deltaMp}
\end{equation}
where $m_{high,0}$ is the maximum between the medians of the default and high-mass posterior, $m_{final,0}$ is the median of the \final posterior, and $m_{final,+\sigma}$ is the $0.84$ quantile of the \final posterior. $\Delta_{M+}=0$ if $m_{high,0}<m_{final,0}$.
\begin{equation}
\Delta_{M-} = \frac{ m_{final,0}-m_{low,0}}{m_{final,0}-m_{final,-\sigma}},
\end{equation}
where $m_{low,0}$ is the minimum of the \default and \highm posterior medians, and $m_{final,-\sigma}$ is the $0.16$ quantile of the final posterior. $\Delta_{M-}=0$ if $m_{low,0}>m_{final,0}$. The values of $\Delta_{M}$ obtained for each planet is given in Tables \ref{tab:planet1} and \ref{tab:planet2}, and discussed in sec. \ref{sec:massdeg}.

Transit timing variations can be studied by pre-extracting all the transit timings of the planets, then fitting these transit timings. Alternatively, one can analyse TTVs using a photo-dynamical model \citep{RaHo2010}; in which the ideal light-curve, accounting for TTVs, is modelled and then fit to the data. \cite{Leleu2023} showed that photo-dynamical analysis often allows for a more robust mass estimate, especially for systems harbouring planets whose individual transit signal-to-noise ratio (S/N) is low (typically $\lesssim 3.5$). We therefore performed a photo-dynamical analysis of the data.


\subsection{Data modelling}
\label{sec:data_modelling}
We performed a joint analysis of all available transit events as well as {the} RV {data}. The gravitational interaction of the 5 resonant planets ($c$ to $g$) {was} also taken into account. For these planets, we used wide, flat priors for the mean longitude, period, impact parameter, and ratio of the radius of the planet over the radius of the star, $R_p / R_\star$. The mass, $e \cos \varpi$ and $e \sin \varpi$ priors were given in section \ref{sec:approach}. The stellar density and stellar radius have Gaussian priors with the values given in Table \ref{tab:star}.

\subsubsection{Photometry}

The photo-dynamical model of the planetary signals, presented in detail in \cite{RIVERS1,Leleu2023}, is computed by predicting transit timings using the TTVfast package \citep{DeAgHo2014} for the outer five planets, and a circular orbit for TOI-178b, and modeling the transits of each planet using the batman package \cite{batman}.
We model the instrumental systematics as well as stellar activity using a linear combination of indicators and B-splines.
We use B-splines as functions of time for all photometric data, 
and an additional B-spline is added with respect to the roll-angle for each CHEOPS visit.
The temporal B-splines have nodes separated by 0.4 days for all photometry, and the periodic B-splines on the roll angle for CHEOPS have nodes separated by 0.1 radian.
For CHEOPS, we used as indicators the telescope tube temperature, the position of the centroid along the Y direction, and the background contamination.
For NGTS we used the airmass.
For TESS, the light curves were pre-detrended so no indicator were used.
This model (indicators and B-splines) only introduces linear parameters.
To accelerate the exploration of the parameter space, we compute the likelihood marginalized over these linear parameters, using the linmarg\footnote{\url{https://gitlab.unige.ch/delisle/linmarg}} python package \citep{Leleu2023}.
The limb-darkening parameters have Gaussian priors computed by LDCU \citep{Deline2022} for each photometric instrument. In addition, a jitter term is added per instrument with a flat prior.

\subsubsection{Radial velocities}
\label{sec:dataRV}

The RV model of the planetary signals is also computed using the TTVfast package \citep{DeckAgol2016} for the 5 outer planets, and a circular model for TOI-178b. We fitted these data by modeling the planets' orbits as well as stellar activity using a Gaussian process (GP) model trained simultaneously on the RVs and ancillary activity indicators using the spleaf\footnote{\url{https://gitlab.unige.ch/delisle/spleaf}} package with 
the FENRIR 2 modes - 4 harmonics - Matérn 1/2 kernel, see appendix \ref{ap:RV} and \cite{Hara2023}. The GP is trained simultaneously on the RVs, the FWHM, and the $H\alpha$ activity indicators.




\section{Results of the data analysis}
\label{sec:results}

\begin{figure}[!ht]
\begin{center}
\includegraphics[width=0.99\linewidth]{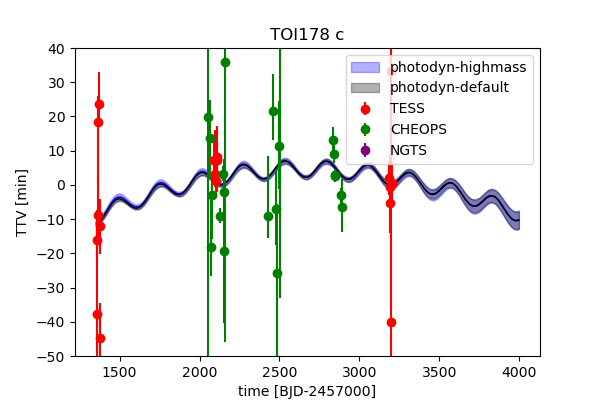}\\
\includegraphics[width=0.99\linewidth]{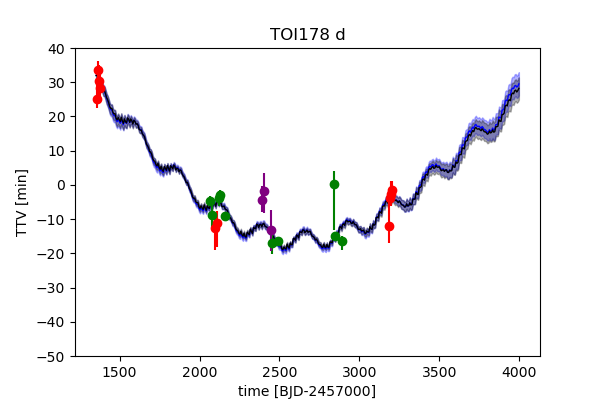}\\
\includegraphics[width=0.99\linewidth]{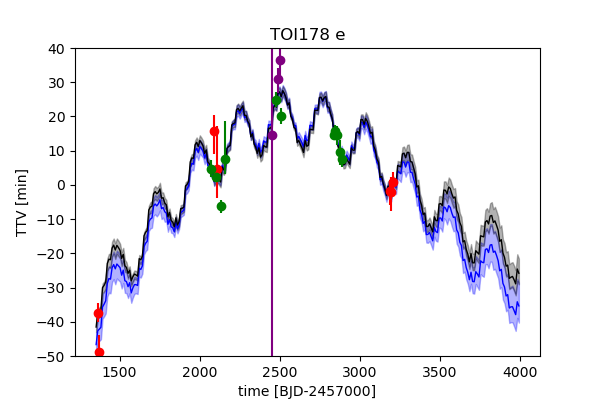}
\caption{\label{fig:TTVcde} Transit timing variations observed for TOI178\,c, $d$ and e. The filled area correspond to the 1$\sigma$ posterior of the photo-dynamical fits presented in section \ref{sec:approach}. The error bars show an estimation of the 1$\sigma$ interval for the mid-transit timing of individual transits for each instrument. See the text for more details.
}
\end{center}
\end{figure}

\begin{figure}[!ht]
\begin{center}
\includegraphics[width=0.99\linewidth]{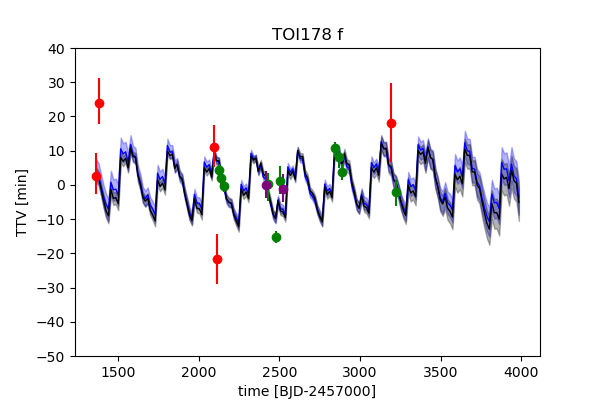}\\
\includegraphics[width=0.99\linewidth]{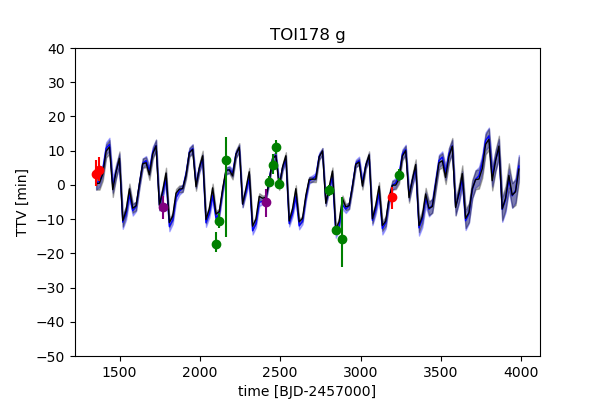}
\caption{\label{fig:TTVfg} Transit timing variations observed for TOI178\,f and g. See Fig. \ref{fig:TTVcde}.
}
\end{center}
\end{figure}

The transit timings estimated for each planet, as well as 300 samples of the final posterior, can be found online. In particular, the transit timings are propagated until 2030, to be usable for follow-up observations of the system.

\subsection{Recovered TTV signal}

Figures \ref{fig:TTVcde} and \ref{fig:TTVfg} show the TTVs of the full fit model for both the \textit{default} and \textit{highmass} set of priors, along with estimations of individual transit timings for each instrument. In these TTV plots, in addition to the super-period at $\sim$260 days visible for all planets, the long-term evolution of the resonant angles start to be visible for planets $c$, $d$ and $e$. The saw-tooth shape of the chopping effect \citep{Deck2015} can be seen on the posterior of planets $f$ and $g$. 

We also fit the individual transit timings to illustrate how each portion of the TTVs curve is constrained. These individual transit timings were estimated by an additional fit, using the full model described in section \ref{sec:data_ana}, but adding each transit timing as a free parameter instead of using the timings of the n-body integration. 
These individual transit timings are shown as colored circles in Fig. \ref{fig:TTVcde} and \ref{fig:TTVfg}. While individual transit timings follow the TTV pattern recovered by the photo-dynamical fit for planets $d$ to $g$. For planet $c$ the median error of the transit timings is larger than the amplitude of the TTV signal. The median standard deviation of the timing derived from the photo-dynamical model ($0.86$ minutes) highlights the necessity of using photo-dynamical modelling for planets with low-S/N transits \citep{Leleu2023}.   

A downside of the photo-dynamical analysis is its model dependency: we have to choose \textit{a priori} how many planets we consider to model the light curves, and the effect of potential missed planets can be hard to identify on the light curve residuals.
Having the individual transit timings also allows us to check if there are hints of additional planets in the resonant chain, for example at orbital periods larger than $P_g=20.7$\,days. Indeed, the effect of this additional planet, which is not modelled by the 5-resonant planet model, could show when subtracting the timing from the photo-dynamical analysis to the individual transit timings. However, beside a couple of strong outliers in the TESS timing of TOI-178 $f$, we did not find strong evidence of a planet significantly impacting the TTVs of the observed planets. We analysed the TESS SAP and PDCSAP 2-minute light curves, as well as background and centroid time series to attempt to understand whether systematic noise on the TESS detector could be causing the two anomalously timed transits of TOI-178 $f$ at 1380.16 [BJD-2457000] and
 2111.26 [BJD-2457000]. We found no sign of systematic noise in the $\sim$12hrs around transit. These outliers might be due to the relatively low SNR of the transits in the TESS data combined with an under estimation of the error bars.

\subsection{Recovered planetary parameters}

\begin{table*} 
\begin{small} 
\caption{Fitted and derived properties of the planets of TOI-178} 
\label{tab:planet1} 
\centering 
\begin{tabular}{llll} 
\hline 
Parameter &  Prior &  photo-dynamical & photo-dynamical + RV\\ 
\hline\hline 
\multicolumn{4}{c}{TOI-178b}\\ 
\hline 
$t_0$ [day]&$\mathcal{U}$(-1.0e+30,1.0e+30)&$1931.17893_{-6.6e-04}^{+7.7e-04}$&$1931.1793_{-0.0010}^{+0.0012}$ \\ 
$P$ [day]&$\mathcal{U}$(0.0e+00,1.0e+04)&$1.9145603_{-2.3e-06}^{+3.9e-06}$&$1.9145601_{-2.9e-06}^{+4.4e-06}$ \\ 
$M/M_\sun$& $^*$ &$2.2e-06_{-2.0e-06}^{+1.3e-05}$&$2.9e-06_{-2.0e-06}^{+2.1e-06}$ \\ 
$R/R_\star$&$\mathcal{U}$(0.0e+00,2.0e-01)&$0.01691_{-6.3e-04}^{+4.4e-04}$&$0.01663_{-4.9e-04}^{+5.3e-04}$ \\ 
$b$&$\mathcal{U}$(0.0e+00,1.5e+00)&$0.417_{-0.039}^{+0.006}$&$0.336_{-0.095}^{+0.081}$ \\ 
dF [ppm]& derived &$285.9_{-20.9}^{+15.1}$&$276.6_{-15.9}^{+17.8}$ \\ 
$M[M_\text{Earth}]$& derived & - &$0.96_{-0.65}^{+0.70}$ \\ 
$R[R_\text{Earth}]$& derived &$1.227_{-0.046}^{+0.034}$&$1.200_{-0.037}^{+0.041}$ \\ 
$\rho[\rho_\text{Earth}]$& derived & - &$0.57_{-0.39}^{+0.40}$ \\ 
\hline\hline 
\multicolumn{4}{c}{TOI-178c}\\ 
\hline 
$\lambda$ [deg]&$\mathcal{U}$(-360.00,360.00)&$54.895_{-0.012}^{+0.015}$&$54.85_{-0.13}^{+0.15}$ \\ 
$P$ [day]&$\mathcal{U}$(3.14,3.34)&$3.2384871_{-7.3e-06}^{+8.9e-06}$&$3.238486_{-1.1e-05}^{+1.2e-05}$ \\ 
$e\cos \varpi$ & $^*$ &$2.4e-05_{-2.3e-04}^{+4.4e-04}$&$2.5e-05_{-3.7e-04}^{+5.4e-04}$ \\ 
$e\sin \varpi$ & $^*$ &$8.5e-06_{-3.0e-04}^{+5.8e-04}$&$3.5e-05_{-3.8e-04}^{+7.9e-04}$ \\ 
$M/M_\sun$ & $^*$ &$1.5e-05_{-2.6e-06}^{+2.5e-06}$&$1.4e-05_{-1.6e-06}^{+1.5e-06}$ \\ 
$R/R_\star$ &$\mathcal{U}$(0.0e+00,2.0e-01)&$0.02447_{-4.6e-04}^{+4.2e-04}$&$0.02431_{-4.4e-04}^{+3.8e-04}$ \\ 
$b$&$\mathcal{U}$(0.0e+00,1.5e+00)&$0.422_{-0.008}^{+0.014}$&$0.394_{-0.042}^{+0.039}$ \\ 
dF [ppm]& derived &$599.0_{-22.3}^{+20.9}$&$590.8_{-21.0}^{+18.7}$ \\ 
$e$& derived &$3.6e-04_{-3.0e-04}^{+6.6e-04}$&$3.2e-04_{-2.0e-04}^{+6.5e-04}$ \\ 
$\varpi$ [deg]& derived &$15_{-110}^{+96}$&$26_{-128}^{+90}$ \\ 
$M[M_\text{Earth}]$& derived &$5.00_{-0.87}^{+0.83}$&$4.64_{-0.53}^{+0.52}$ \\ 
$R[R_\text{Earth}]$& derived &$1.777_{-0.035}^{+0.034}$&$1.754_{-0.040}^{+0.032}$ \\ 
$\rho[\rho_\text{Earth}]$& derived &$0.91_{-0.18}^{+0.12}$&$0.87_{-0.10}^{+0.11}$ \\ 
$\Delta_M $ & &0.72 & 0.55\\ 
\hline\hline 
\multicolumn{4}{c}{TOI-178d}\\ 
\hline 
$\lambda$ [deg]&$\mathcal{U}$(-360.00,360.00)&$28.525_{-0.013}^{+0.011}$&$28.36_{-0.16}^{+0.15}$ \\ 
$P$ [day]&$\mathcal{U}$(6.36,6.75)&$6.557593_{-8.4e-05}^{+9.0e-05}$&$6.557569_{-5.6e-05}^{+6.5e-05}$ \\ 
$e\cos \varpi$ & $^*$ &$-0.00637_{-4.2e-04}^{+5.0e-04}$&$-0.0052_{-0.0014}^{+0.0012}$ \\ 
$e\sin \varpi$ & $^*$ &$-0.0047_{-0.0020}^{+0.0025}$&$-0.0028_{-0.0026}^{+0.0021}$ \\ 
$M/M_\sun$ & $^*$ &$1.7e-05_{-1.2e-06}^{+1.1e-06}$&$1.6e-05_{-1.3e-06}^{+1.2e-06}$ \\ 
$R/R_\star$ &$\mathcal{U}$(0.0e+00,2.0e-01)&$0.03755_{-3.5e-04}^{+3.7e-04}$&$0.03737_{-4.1e-04}^{+3.5e-04}$ \\ 
$b$&$\mathcal{U}$(0.0e+00,1.5e+00)&$0.511_{-0.004}^{+0.012}$&$0.501_{-0.027}^{+0.023}$ \\ 
dF [ppm]& derived &$1409.8_{-26.3}^{+28.0}$&$1396.4_{-30.6}^{+26.6}$ \\ 
$e$& derived &$0.0080_{-0.0014}^{+0.0015}$&$0.0068_{-0.0016}^{+0.0016}$ \\ 
$\varpi$ [deg]& derived &$-143.8_{-16.5}^{+11.0}$&$-146.3_{-15.6}^{+20.8}$ \\ 
$M[M_\text{Earth}]$& derived &$5.70_{-0.40}^{+0.36}$&$5.20_{-0.43}^{+0.39}$ \\ 
$R[R_\text{Earth}]$& derived &$2.728_{-0.037}^{+0.033}$&$2.695_{-0.046}^{+0.041}$ \\ 
$\rho[\rho_\text{Earth}]$& derived &$0.280_{-0.020}^{+0.023}$&$0.265_{-0.024}^{+0.024}$ \\ 
$\Delta_M $ & &0.84 & 0.52\\ 
\end{tabular} 
\tablefoot{  
results of the photo-dynamical and photo-dynamical+RV fits for the \textit{final} set of priors. The orbital elements are given at the date 2458352.55018382 BJD. $\lambda$ is the mean longitude of the planet, $\varpi$ its longitude of periastron. $\Delta_M$ is the robustness criterion defined in equation (\ref{eq:deltaM}). $b$, $e$, $M$ and $R$ are the planet's impact parameter, eccentricity, mass and radius respectively. $M_\odot$ is the mass of the sun and $R_\star$ is the radius of the star. $^*$ The mass and eccentricity priors depend on the case, see section \ref{sec:approach}.}
\end{small} 
\end{table*}
\begin{table*} 
\begin{small} 
\caption{Fitted and derived properties of the planets of TOI-178} 
\label{tab:planet2} 
\centering 
\begin{tabular}{llll} 
\hline 
Parameter &  Prior & photo-dynamical & photo-dynamical + RV  \\ 
\hline\hline 
\multicolumn{4}{c}{TOI-178e}\\ 
\hline 
$\lambda$ [deg]&$\mathcal{U}$(-360.00,360.00)&$74.265_{-0.017}^{+0.017}$&$74.23_{-0.11}^{+0.11}$ \\ 
$P$ [day]&$\mathcal{U}$(9.66,10.26)&$9.96336_{-1.5e-04}^{+1.6e-04}$&$9.96318_{-1.2e-04}^{+1.5e-04}$ \\ 
$e\cos \varpi$ & $^*$ &$-9.8e-05_{-5.1e-04}^{+1.7e-04}$&$-2.0e-05_{-6.7e-04}^{+4.3e-04}$ \\ 
$e\sin \varpi$ & $^*$ &$-2.7e-05_{-5.2e-04}^{+3.5e-04}$&$1.0e-07_{-5.7e-04}^{+5.7e-04}$ \\ 
$M/M_\sun$ & $^*$ &$1.0e-05_{-1.0e-06}^{+1.1e-06}$&$1.0e-05_{-8.6e-07}^{+8.6e-07}$ \\ 
$R/R_\star$ &$\mathcal{U}$(0.0e+00,2.0e-01)&$0.03205_{-4.7e-04}^{+4.5e-04}$&$0.03191_{-3.7e-04}^{+3.8e-04}$ \\ 
$b$&$\mathcal{U}$(0.0e+00,1.5e+00)&$0.5842_{-0.0073}^{+0.0099}$&$0.574_{-0.027}^{+0.015}$ \\ 
dF [ppm]& derived &$1026.9_{-30.0}^{+28.9}$&$1018.1_{-23.5}^{+24.7}$ \\ 
$e$& derived &$4.3e-04_{-3.6e-04}^{+7.5e-04}$&$3.8e-04_{-2.5e-04}^{+6.7e-04}$ \\ 
$\varpi$ [deg]& derived &$-49_{-95}^{+175}$&$2_{-134}^{+120}$ \\ 
$M[M_\text{Earth}]$& derived &$3.34_{-0.33}^{+0.37}$&$3.48_{-0.29}^{+0.29}$ \\ 
$R[R_\text{Earth}]$& derived &$2.325_{-0.038}^{+0.034}$&$2.301_{-0.039}^{+0.038}$ \\ 
$\rho[\rho_\text{Earth}]$& derived &$0.264_{-0.023}^{+0.029}$&$0.287_{-0.026}^{+0.028}$ \\ 
$\Delta_M $ & &0.54 & 0.49\\ 
\hline\hline 
\multicolumn{4}{c}{TOI-178f}\\ 
\hline 
$\lambda$ [deg]&$\mathcal{U}$(-360.00,360.00)&$157.735_{-0.005}^{+0.021}$&$157.67_{-0.09}^{+0.11}$ \\ 
$P$ [day]&$\mathcal{U}$(14.78,15.69)&$15.23351_{-1.6e-04}^{+1.4e-04}$&$15.23335_{-2.5e-04}^{+2.4e-04}$ \\ 
$e\cos \varpi$ & $^*$ &$-2.1e-04_{-4.6e-04}^{+2.9e-04}$&$-8.4e-06_{-6.7e-04}^{+8.1e-04}$ \\ 
$e\sin \varpi$ & $^*$ &$8.8e-04_{-0.0009}^{+0.0015}$&$0.0017_{-0.0016}^{+0.0016}$ \\ 
$M/M_\sun$ & $^*$ &$1.9e-05_{-1.5e-06}^{+1.2e-06}$&$1.7e-05_{-1.2e-06}^{+1.4e-06}$ \\ 
$R/R_\star$ &$\mathcal{U}$(0.0e+00,2.0e-01)&$0.03370_{-5.9e-04}^{+5.7e-04}$&$0.03351_{-6.0e-04}^{+5.6e-04}$ \\ 
$b$&$\mathcal{U}$(0.0e+00,1.5e+00)&$0.7399_{-0.0042}^{+0.0098}$&$0.734_{-0.014}^{+0.015}$ \\ 
dF [ppm]& derived &$1135.6_{-39.3}^{+38.6}$&$1123.1_{-40.0}^{+38.0}$ \\ 
$e$& derived &$0.0011_{-0.0009}^{+0.0014}$&$4.5e-04_{-2.6e-04}^{+7.0e-04}$ \\ 
$\varpi$ [deg]& derived &$98.2_{-73.8}^{+28.8}$&$87.3_{-40.1}^{+28.5}$ \\ 
$M[M_\text{Earth}]$& derived &$6.24_{-0.49}^{+0.40}$&$5.63_{-0.41}^{+0.45}$ \\ 
$R[R_\text{Earth}]$& derived &$2.443_{-0.053}^{+0.047}$&$2.417_{-0.048}^{+0.041}$ \\ 
$\rho[\rho_\text{Earth}]$& derived &$0.426_{-0.037}^{+0.040}$&$0.402_{-0.036}^{+0.033}$ \\ 
$\Delta_M $ & &0.93 & 1.15\\ 
\hline\hline 
\multicolumn{4}{c}{TOI-178g}\\ 
\hline 
$\lambda$ [deg]&$\mathcal{U}$(-360.00,360.00)&$55.329_{-0.019}^{+0.007}$&$55.27_{-0.10}^{+0.06}$ \\ 
$P$ [day]&$\mathcal{U}$(20.09,21.34)&$20.71700_{-3.4e-04}^{+4.1e-04}$&$20.71663_{-3.0e-04}^{+3.1e-04}$ \\ 
$e\cos \varpi$ & $^*$ &$-2.0e-04_{-2.9e-04}^{+2.1e-04}$&$1.2e-04_{-4.2e-04}^{+7.9e-04}$ \\ 
$e\sin \varpi$ & $^*$ &$-9.3e-05_{-6.9e-04}^{+2.6e-04}$&$-1.7e-04_{-0.0012}^{+0.0005}$ \\ 
$M/M_\sun$ & $^*$ &$1.4e-05_{-7.9e-07}^{+8.8e-07}$&$1.3e-05_{-1.1e-06}^{+1.2e-06}$ \\ 
$R/R_\star$ &$\mathcal{U}$(0.0e+00,2.0e-01)&$0.04086_{-5.4e-04}^{+4.4e-04}$&$0.04075_{-4.4e-04}^{+5.1e-04}$ \\ 
$b$&$\mathcal{U}$(0.0e+00,1.5e+00)&$0.8591_{-0.0029}^{+0.0027}$&$0.8549_{-0.0054}^{+0.0055}$ \\ 
dF [ppm]& derived &$1669.2_{-43.7}^{+36.0}$&$1660.5_{-35.5}^{+41.8}$ \\ 
$e$& derived &$4.5e-04_{-3.5e-04}^{+5.4e-04}$&$4.3e-04_{-2.8e-04}^{+6.5e-04}$ \\ 
$\varpi$ [deg]& derived &$-98_{-44}^{+227}$&$-44_{-63}^{+120}$ \\ 
$M[M_\text{Earth}]$& derived &$4.58_{-0.26}^{+0.29}$&$4.40_{-0.37}^{+0.39}$ \\ 
$R[R_\text{Earth}]$& derived &$2.964_{-0.048}^{+0.042}$&$2.939_{-0.055}^{+0.057}$ \\ 
$\rho[\rho_\text{Earth}]$& derived &$0.177_{-0.012}^{+0.013}$&$0.174_{-0.015}^{+0.015}$ \\ 
$\Delta_M $ & &1.34 & 0.79\\ 
\end{tabular} 
\tablefoot{  
results of the photo-dynamical and photo-dynamical+RV fits for the \textit{final} set of priors. The orbital elements are given at the date 2458352.55018382 BJD. $\lambda$ is the mean longitude of the planet, $\varpi$ its longitude of periastron. $\Delta_M$ is the robustness criterion defined in equation (\ref{eq:deltaM}). $b$, $e$, $M$ and $R$ are the planet's impact parameter, eccentricity, mass and radius respectively. $M_\odot$ is the mass of the sun and $R_\star$ is the radius of the star. $^*$ The mass and eccentricity priors depend on the case, see section \ref{sec:approach}.}
\end{small} 
\end{table*}


\begin{figure}[!ht]
\begin{center}
\includegraphics[width=0.99\linewidth]{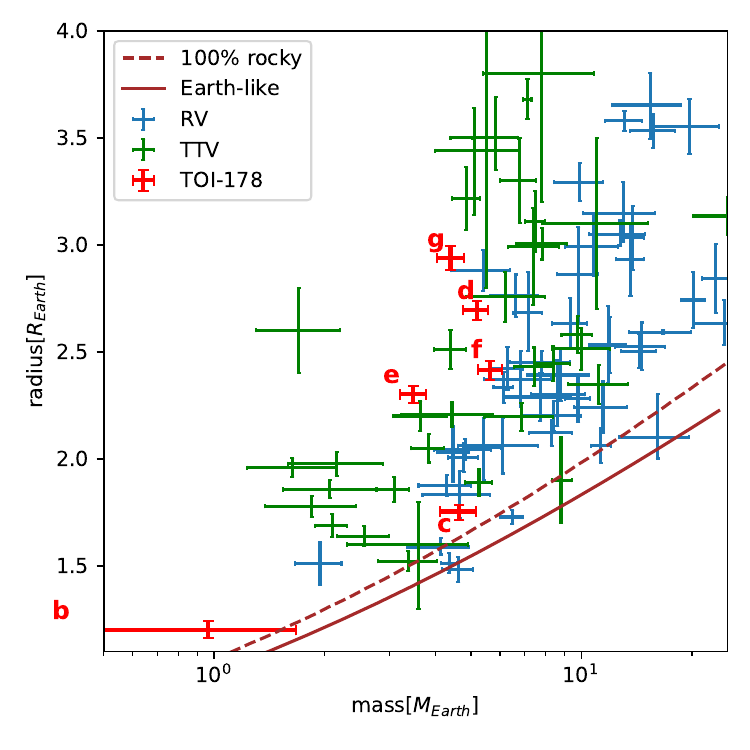}
\caption{\label{fig:MR} Mass-radius relationship of the TOI-178 planets (final posterior), compared with the super-Earth to Mini-Neptune populations for which the mass was derived through RV \citep{Otegi2020} and TTVs \citep{HaLi2017,Leleu2023}. {The composition lines `terrestrial Earth-like', in solid brown, and pure MgSiO$_3$ `rocky', in dashed brown, are taken from \citet{Zeng2016}.}
}
\end{center}
\end{figure}
\begin{figure}[!ht]
\begin{center}
\includegraphics[width=0.99\linewidth]{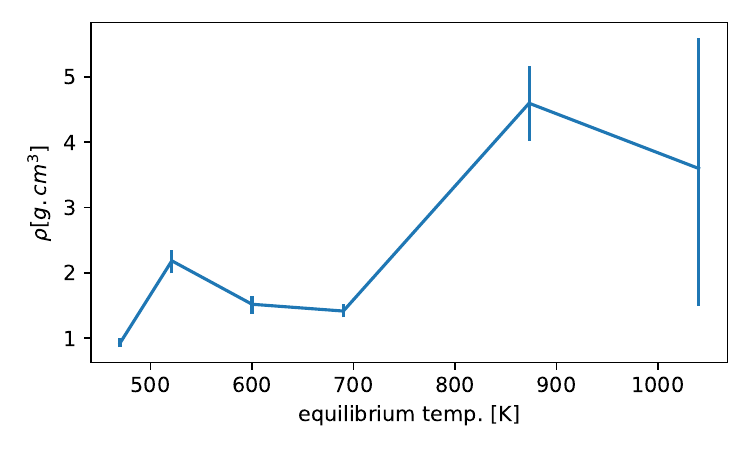}
\caption{\label{fig:Teq_density} Mean density of the TOI-178 planets as a function of their equilibrium temperature.
}
\end{center}
\end{figure}

The planetary masses, radii, transit parameters and Jacobi orbital elements at epoch 1352.55018 [BJD-2457000] are given in tables \ref{tab:planet1} and \ref{tab:planet2} for the \textit{final} posterior of the photo-dynamical fit and the combined photo-dynamical and radial-velocity fits, while Table \ref{tab:stellarParam} shows the fitted stellar and noise parameters. 

In this joint analysis, the final posterior reaches a precision of 12\% for the mass of planet $c$, while the precision on the mass of planets $d$ to $g$ is better than $10\%$. The precision we reach for planet $b$ remains similar to the one published in L21 since the planet does not induce any significant TTVs on the other planets of the system. The precision on the planetary radius ranges from $\approx 3\%$ for planet $b$ to better than $2 \%$ for the largest planet of the system. 
Figure \ref{fig:MR} shows the newly-determined mass-radius relation of TOI-178 planets with respect to the planets whose mass was characterised by either TTVs (green) or RV (blue). These reference planets were chosen for their robustness to the choice of prior for the TTV population and for their precision for the RV population. The RV-characterised planets appear to be denser than the TTV-characterised ones, although we note that selecting RV-characterised planets based on their precision might bias that population toward denser mass-radius {combinations}.
The apparent discrepancy between the TTV and RV-characterised planets has been heavily discussed
\citep[e.g.][]{WuLi2013,WeissMarcy2014,MillsMazeh2017,HaLi2017,Cubillos2017,Millholland2020,Leleu2023,Adibekyan2024}. In particular, \cite{HaLi2017} put forward as a possible explanation a selection bias, since TTVs tend to allow the characterisation of small planets on larger orbital periods, hence cooler orbits, than the bulk of the RV characterisation. It has also been proposed that the systems formed in different environments, such as a different amount of available iron during the planets' formation \citep{Adibekyan2024}. 
The gaseous planets of TOI-178 appear to be on the lower end of density for their respective radius, more akin to the sub-population that was characterised using TTVs. The possibility to characterise TOI-178 both by RV and TTVs makes this system especially relevant to study this discrepancy. With the preliminary masses obtained by the RV data, L21 showed that the density of the planets appeared to not evolve {monotonically} with respect to the equilibrium temperature of the planets. With our updated planetary masses and radii, we retrieve a similar trend, see Fig. \ref{fig:Teq_density}.

\begin{figure}[!ht]
\begin{center}
\includegraphics[width=0.99\linewidth]{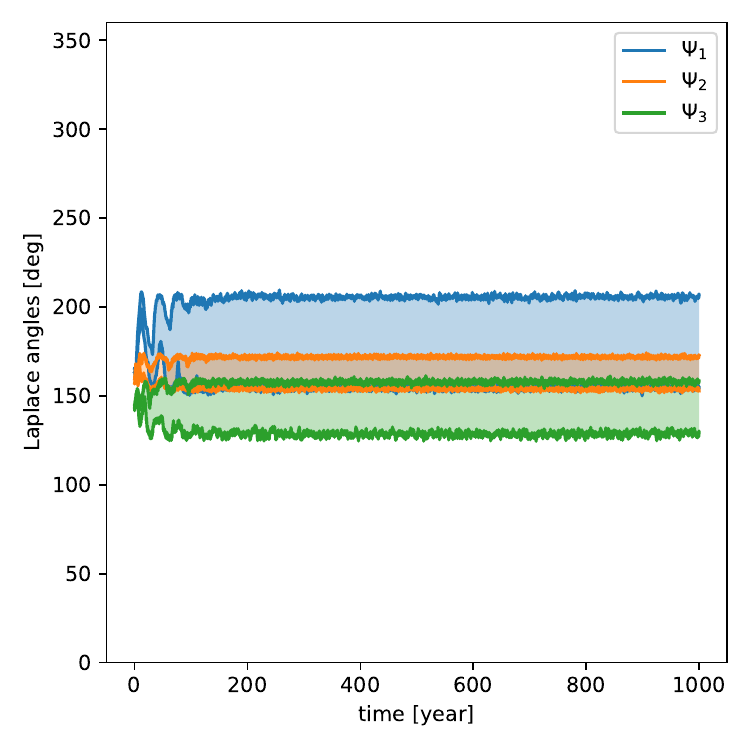}\\
\caption{\label{fig:Laplace_angles} 3-sigma envelope of the evolution of the Laplace angles $\Psi_1$, $\Psi_2$ and $\Psi_3$ over 1000 years for the final posterior.
}
\end{center}
\end{figure}

Finally, L21 claimed that the system is in a chain of Laplace resonances. Using the preliminary mass determination, averaged periods and assuming zero initial eccentricities, they showed that the system was within the stability domain of the 3-body resonance chain. Thanks to the joint photo-dynamical and RV fit, we now have estimations of the instantaneous orbital elements of each planet at a given date, see Tables \ref{tab:planet1} and \ref{tab:planet2}. Using 300 randomly-selected samples of the final posterior, we can simulate the future evolution of the Laplace angles $ \psi_1=1\lambda_c-4\lambda_d+3\lambda_e$, 
$ \psi_2=2\lambda_d-5\lambda_e+3\lambda_f$, and $ \psi_3= 1\lambda_e-3\lambda_f+2\lambda_g$, where $\lambda$ is the mean longitude of the planet. Figure \ref{fig:Laplace_angles} shows the 3-sigma envelope of the evolution of these angles over the next 1000 years. Although we lose the information on the phase of the {libration} after a few decades, we can see that our full posterior yields librating angles, confirming that the system is indeed currently librating in the chain of Laplace resonances, around the equilibrium described in Table 6 of L21.

\subsection{Robustness of retrieved masses}
\label{sec:massdeg}

\begin{figure*}[!ht]
\begin{center}
\includegraphics[width=0.80\linewidth]{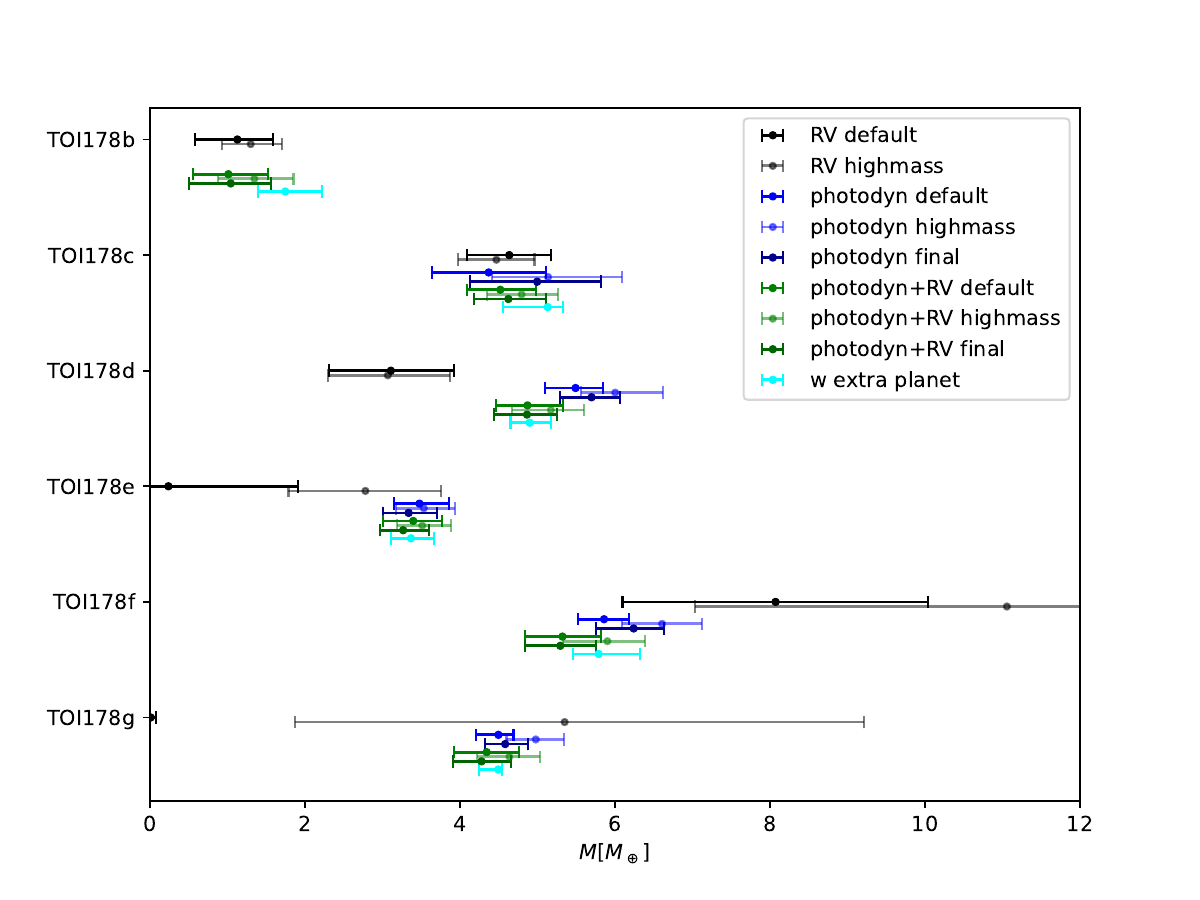}
\caption{\label{fig:masspost} Mass determination for different datasets and choices of priors. In black are the posterior from the RV analysis with the FENRIR 2 modes - 4 harmonics - Matérn 1/2 model, see appendix \ref{ap:RV}. The blue posteriors are from the photo-dynamical model alone, while the green posteriors are from the joint photo-dynamical and RV fits. The default, highmass, and final set of priors are defined in section \ref{sec:approach}. The cyan posterior synthesise the best fits of all our models with additional planets in the resonant chain, see appendix \ref{ap:added_planet}. The photodyn \textit{final} and photodyn+RV \textit{final} are the cases reported in Tables \ref{tab:planet1} and \ref{tab:planet2}.
}
\end{center}
\end{figure*}

To estimate the robustness of mass determination with each technique, we performed RV fits, photo-dynamical fits, and joint fits with the sets of priors defined in section \ref{sec:approach}. Note that for the RV fit the eccentricity of the planets is set to zero and therefore the \textit{default} and \textit{final} sets of prior are identical. 
For the RV fits, we additionally tested several activity models (see Appendix~\ref{ap:RV}),
as it might have an impact on the mass determination 
\citep[e.g.,][]{Bonfils2018,Ahrer2021}.



%
%
We show in Fig.~\ref{fig:masspost}, a summary of the mass posteriors from these analyses.
We see in Fig.~\ref{fig:masspost} that the RV mass determination (black and gray) is robust for the three inner planets,
in the sense that the posteriors for the default and the high-mass priors agree well.
The three outer planets' masses are more sensitive to the priors.
We also note that these planets' masses are also more sensitive to a change of stellar activity model (see Appendix~\ref{ap:RV}).
We recall that the stellar rotation period was estimated to be about 35~d by L21, and that we only have 46 RV measurements over 113 days.
It is thus very challenging to train a GP at the same time as fitting for the 6 planets' masses on these data.
Since we noticed a strong model sensitivity for the 3 outer planets' mass, we selected one of the activity models that exhibited large uncertainties for these 3 parameters for the subsequent analyses (joint fit), in order to avoid biasing the final mass estimates for these planets (see Appendix~\ref{ap:RV}).
This does not mean that the chosen activity model is intrinsically better,
and better constraining the activity model and the planets' masses from RV would require more measurements.

%



The posteriors of the masses from the photo-dynamical fit for the different sets of priors are shown in blue in Fig.  \ref{fig:masspost}. As explained in section \ref{sec:approach}, the $\Delta_M$ criterion quantifies the prior-dependency of these masses. $\Delta_M=0$ indicates that the posteriors of the default, highmass and final fit all perfectly agree, while $\Delta_M=1$ indicates that the median of the default or highmass posteriors lies exactly at $1-\sigma$ from the median of the final posterior. \cite{Leleu2023} set an arbitrary limit at $\Delta_M=1.3$ to estimate if the retrieved mass was degenerate or not. From the photo-dynamical fit alone, planets $c$, $d$, $e$ and $f$ have robust masses, with values of $\Delta_M=0.72$, $0.84$, $0.54$ and $0.93$, while the mass of planet $g$ shows a larger prior dependency, with $\Delta_M=1.34$ (see Tables \ref{tab:planet1} and \ref{tab:planet2}).


Overall, the photo-dynamical model gives better mass precision than the RV fit for planets $d$ to $e$, while having a much lower prior dependency for the three outer planets. The mass of TOI-178c is fully consistent between the two techniques, which is important to note, since this mass is also the least sensitive to the choice of activity model or mass prior in the RV analysis, see appendix \ref{ap:RV}. For TOI-178d the median of the \textit{default} photo-dynamic posterior is 3-$\sigma$ away from the RV \textit{default} posterior. However, while the posteriors shown in Fig. \ref{fig:masspost} display a low dependency on the mass prior, the dependency to the choice of activity model is bigger, see Appendix~\ref{ap:RV}. 

When looking at the posteriors of the joint fit, in green in Fig \ref{fig:masspost}, we can see that most of the constraints on the mass come from the photo-dynamical model, as the joint mass posteriors (green) remain close to the photo-dynamical posteriors (blue). We nonetheless note a general improvement in the robustness of the determined mass, as the prior dependency ($\Delta_M$ in Tables \ref{tab:planet1} and \ref{tab:planet2}) is generally smaller in the joint fit than in the photo-dynamical fit alone, beside a slight increase to 1.15 for TOI-178f. 

Finally, we estimate the robustness of the derived masses to the existence of a 7th planet in the resonant chain, see appendix \ref{ap:added_planet} for more detail. The outcome of this test is reported by the cyan error bars in Fig. \ref{fig:masspost}. Overall, adding an additional planet to the model does not change significantly the derived mass for the six known planets. These results are in agreement with the findings of \cite{Leleu2023}, i.e., that having the proper TTV model (number of planets in the system) goes {hand in hand} with having a low prior dependency (low value of $\Delta_M$) for the retrieved masses of the system.

\section{Summary and conclusion}
\label{sec:conclusion}

In this study we present a joint analysis of all available data for the TOI-178 system. We combine a photo-dynamical analysis of the data taken by TESS, CHEOPS, and NGTS, with the RVs taken by ESPRESSO.
In addition, we explore the robustness of the derived masses by comparing the results of the joint fit to the mass constraints obtained by RV and photometry independently.

We show that the available RVs on their own {do not enable us} to fully distinguish the planetary signal from stellar activity for the three outer planets. In particular, for planet $g$ fitting the RV data with different activity models and mass priors results in a large range of unphysical mass measurements, an issue that was also reported for example by \cite{Lopez2019} in the case of K2-138f. 
Arguably, the case of TOI-178d is the most intriguing, with all activity models and choice of priors giving relatively consistent masses using the RV data alone, and these masses {being} at least 2-$\sigma$ away from the {results} of the photo-dynamical fit. On the other hand, the techniques fully agree for the mass of TOI-178c, which is reassuring since it is the mass that is the most robust to the choice of activity model and mass prior across all our tests (see appendix \ref{ap:RV}).

We show that the photo-dynamical analysis is able to retrieve robust masses on its own (low dependency on the chosen set of priors) for planets $c$ to $f$, and our joint photo-dynamical+RV analysis is able to retrieve robust masses for planets $c$ to $g$.
This highlights the need for thorough follow-up campaigns for the characterisation of multi-planetary systems, as well as the development and calibration of robustness criteria for the derived planetary masses. For these criteria, bright resonant systems such as TOI-178 have an important role to play since its planetary masses can be obtained independently by both RV and TTVs. Obtaining these two independent mass measurements for the same planets can ensure the accuracy of these measurements and therefore help {with} the calibration of the robustness criteria.
In the case of TOI-178, additional transit measurements would further increase our confidence in the robustness of the TTV masses, while additional RV measurements are necessary both to obtain better constraints on planet $b$ which is outside of the chain, and to distinguish the signal of the three outer planets from the stellar activity.
These observations also enable a better understanding of the synergies between photometric and RV measurements on a given system, which will be key to achieve the full potential of the upcoming PLATO mission \citep{PLATO-paper}.


Thanks to the intensive follow-up effort, TOI-178 now has one of the best characterised architectures of resonant chains of sub-Neptunes. We {were able to} show that the system is indeed librating inside the Laplace resonant chain, around the equilibrium that was described in section 6.2 of L21. We also reached a precision of $\sim 12\%$ for the mass of planet $c$, and better than $10\%$ for planets $d$, $e$, $f$ and $g$. The new precisions on the radii range from $\approx 3\%$ for planet $b$ to better than $2 \%$ for planet $g$. We note that we were not able to improve the mass of planet $b$ as it is not interacting with the resonant chain.
 The newly derived mass and radius of $c$ {appear to depart from the 100\% rocky composition that was favoured in L21, implying that a gaseous envelope is necessary to reproduce the observed density}. Planet $d$ to $g$ remain akin to the TTV-characterised Mini-Neptunes (see Fig. \ref{fig:MR}), making of TOI-178 a key system in the study of the apparent mass-radius discrepancy between TTV and RV-characterised population \citep{WuLi2013,WeissMarcy2014,MillsMazeh2017,Cubillos2017,HaLi2017,Millholland2020,Leleu2023,Adibekyan2024}. 
We also retrieve the non-monotonous density variations as function as the distance to the star that was hinted at by L21. 

These new constraints will allow better comparison with models of formation and evolution of planetary systems \citep[e.g.][]{NGPPS1,Izidoro2022}, as well as enabling the exploitation of the JWST transmission spectrum of the system to their full potential.

\bibliographystyle{aa}
\bibliography{biblio}

\begin{acknowledgements}
CHEOPS is an ESA mission in partnership with Switzerland with important contributions to the payload and the ground segment from Austria, Belgium, France, Germany, Hungary, Italy, Portugal, Spain, Sweden, and the United Kingdom. The CHEOPS Consortium would like to gratefully acknowledge the support received by all the agencies, offices, universities, and industries involved. Their flexibility and willingness to explore new approaches were essential to the success of this mission. CHEOPS data analysed in this article will be made available in the CHEOPS mission archive (\url{https://cheops.unige.ch/archive_browser/}). 
This work has been carried out within the framework of the NCCR PlanetS supported by the Swiss National Science Foundation under grants 51NF40\_182901 and 51NF40\_205606. AL acknowledges support of the Swiss National Science Foundation under grant number TMSGI2\_211697. 
The Belgian participation to CHEOPS has been supported by the Belgian Federal Science Policy Office (BELSPO) in the framework of the PRODEX Program, and by the University of Liège through an ARC grant for Concerted Research Actions financed by the Wallonia-Brussels Federation. 
L.D. thanks the Belgian Federal Science Policy Office (BELSPO) for the provision of financial support in the framework of the PRODEX Programme of the European Space Agency (ESA) under contract number 4000142531. 
The contributions at the Mullard Space Science Laboratory by E.M.B. have been supported by STFC through the consolidated grant ST/W001136/1. 
ABr was supported by the SNSA. 
This work has been carried out within the framework of the NCCR PlanetS supported by the Swiss National Science Foundation under grants 51NF40\_182901 and 51NF40\_205606. 
TWi acknowledges support from the UKSA and the University of Warwick. 
ML acknowledges support of the Swiss National Science Foundation under grant number PCEFP2\_194576. 
This project has received funding from the Swiss National Science Foundation for project 200021\_200726. It has also been carried out within the framework of the National Centre of Competence in Research PlanetS supported by the Swiss National Science Foundation under grant 51NF40\_205606. The authors acknowledge the financial support of the SNSF. 
ML and HC acknowledge support of the Swiss National Science Foundation under grant number PCEFP2\_194576. 
MNG is the ESA CHEOPS Project Scientist and Mission Representative, and as such also responsible for the Guest Observers (GO) Programme. MNG does not relay proprietary information between the GO and Guaranteed Time Observation (GTO) Programmes, and does not decide on the definition and target selection of the GTO Programme. 
YAl acknowledges support from the Swiss National Science Foundation (SNSF) under grant 200020\_192038. 
We acknowledge financial support from the Agencia Estatal de Investigación of the Ministerio de Ciencia e Innovación MCIN/AEI/10.13039/501100011033 and the ERDF “A way of making Europe” through projects PID2019-107061GB-C61, PID2019-107061GB-C66, PID2021-125627OB-C31, and PID2021-125627OB-C32, from the Centre of Excellence “Severo Ochoa” award to the Instituto de Astrofísica de Canarias (CEX2019-000920-S), from the Centre of Excellence “María de Maeztu” award to the Institut de Ciències de l’Espai (CEX2020-001058-M), and from the Generalitat de Catalunya/CERCA programme. 
This research was funded in whole or in part by the UKRI, (Grants ST/X001121/1, EP/X027562/1). 
We acknowledge financial support from the Agencia Estatal de Investigación of the Ministerio de Ciencia e Innovación MCIN/AEI/10.13039/501100011033 and the ERDF “A way of making Europe” through projects PID2019-107061GB-C61, PID2019-107061GB-C66, PID2021-125627OB-C31, and PID2021-125627OB-C32, from the Centre of Excellence “Severo Ochoa'' award to the Instituto de Astrofísica de Canarias (CEX2019-000920-S), from the Centre of Excellence “María de Maeztu” award to the Institut de Ciències de l’Espai (CEX2020-001058-M), and from the Generalitat de Catalunya/CERCA programme. 
S.C.C.B. acknowledges support from FCT through FCT contracts nr. IF/01312/2014/CP1215/CT0004. 
The contributions from MPB were carried out within the framework of the National Centre of Competence in Research PlanetS supported by the Swiss National Science Foundation under grants 51NF40\_182901 and 51NF40\_205606. The authors acknowledge the financial support of the SNSF. 
LBo, GBr, VNa, IPa, GPi, RRa, GSc, VSi, and TZi acknowledge support from CHEOPS ASI-INAF agreement n. 2019-29-HH.0. 
C.B. acknowledges support from the Swiss Space Office through the ESA PRODEX program. 
ACC acknowledges support from STFC consolidated grant number ST/V000861/1, and UKSA grant number ST/X002217/1. 
P.E.C. is funded by the Austrian Science Fund (FWF) Erwin Schroedinger Fellowship, program J4595-N. 
This project was supported by the CNES. 
This work was supported by FCT - Funda\c{c}\~{a}o para a Ci\^{e}ncia e a Tecnologia through national funds and by FEDER through COMPETE2020 through the research grants UIDB/04434/2020, UIDP/04434/2020, 2022.06962.PTDC. 
O.D.S.D. is supported in the form of work contract (DL 57/2016/CP1364/CT0004) funded by national funds through FCT. 
B.-O. D. acknowledges support from the Swiss State Secretariat for Education, Research and Innovation (SERI) under contract number MB22.00046. 
MF and CMP gratefully acknowledge the support of the Swedish National Space Agency (DNR 65/19, 174/18). 
DG gratefully acknowledges financial support from the CRT foundation under Grant No. 2018.2323 “Gaseousor rocky? Unveiling the nature of small worlds”. 
E.G. gratefully acknowledges support from the UK Science and Technology Facilities Council (STFC. 
project reference ST/W001047/1). 
M.G. is an F.R.S.-FNRS Senior Research Associate. 
FH is supported by an STFC studentship. 
CHe acknowledges support from the European Union H2020-MSCA-ITN-2019 under Grant Agreement no. 860470 (CHAMELEON). 
KGI is the ESA CHEOPS Project Scientist and is responsible for the ESA CHEOPS Guest Observers Programme. She does not participate in, or contribute to, the definition of the Guaranteed Time Programme of the CHEOPS mission through which observations described in this paper have been taken, nor to any aspect of target selection for the programme. 
JSJ greatfully acknowledges support by FONDECYT grant 1201371 and from the ANID BASAL project FB210003. 
K.W.F.L. was supported by Deutsche Forschungsgemeinschaft grants RA714/14-1 within the DFG Schwerpunkt SPP 1992, Exploring the Diversity of Extrasolar Planets. 
This work was granted access to the HPC resources of MesoPSL financed by the Region Ile de France and the project Equip@Meso (reference ANR-10-EQPX-29-01) of the programme Investissements d'Avenir supervised by the Agence Nationale pour la Recherche. 
PM acknowledges support from STFC research grant number ST/R000638/1. 
This work was also partially supported by a grant from the Simons Foundation (PI Queloz, grant number 327127). 
NCSa acknowledges funding by the European Union (ERC, FIERCE, 101052347). Views and opinions expressed are however those of the author(s) only and do not necessarily reflect those of the European Union or the European Research Council. Neither the European Union nor the granting authority can be held responsible for them. 
A. S. acknowledges support from the Swiss Space Office through the ESA PRODEX program. 
S.G.S. acknowledge support from FCT through FCT contract nr. CEECIND/00826/2018 and POPH/FSE (EC). 
The Portuguese team thanks the Portuguese Space Agency for the provision of financial support in the framework of the PRODEX Programme of the European Space Agency (ESA) under contract number 4000142255. 
GyMSz acknowledges the support of the Hungarian National Research, Development and Innovation Office (NKFIH) grant K-125015, a a PRODEX Experiment Agreement No. 4000137122, the Lendulet LP2018-7/2021 grant of the Hungarian Academy of Science and the support of the city of Szombathely. 
V.V.G. is an F.R.S-FNRS Research Associate. 
JV acknowledges support from the Swiss National Science Foundation (SNSF) under grant PZ00P2\_208945. 
NAW acknowledges UKSA grant ST/R004838/1. 

ACMC acknowledges support from the FCT, Portugal, through the CFisUC projects UIDB/04564/2020 and UIDP/04564/2020, with DOI identifiers 10.54499/UIDB/04564/2020 and 10.54499/UIDP/04564/2020, respectively.
\end{acknowledgements}

\appendix

\section{EulerCam}

\begin{figure}
\begin{center}
\includegraphics[width=0.99\linewidth]{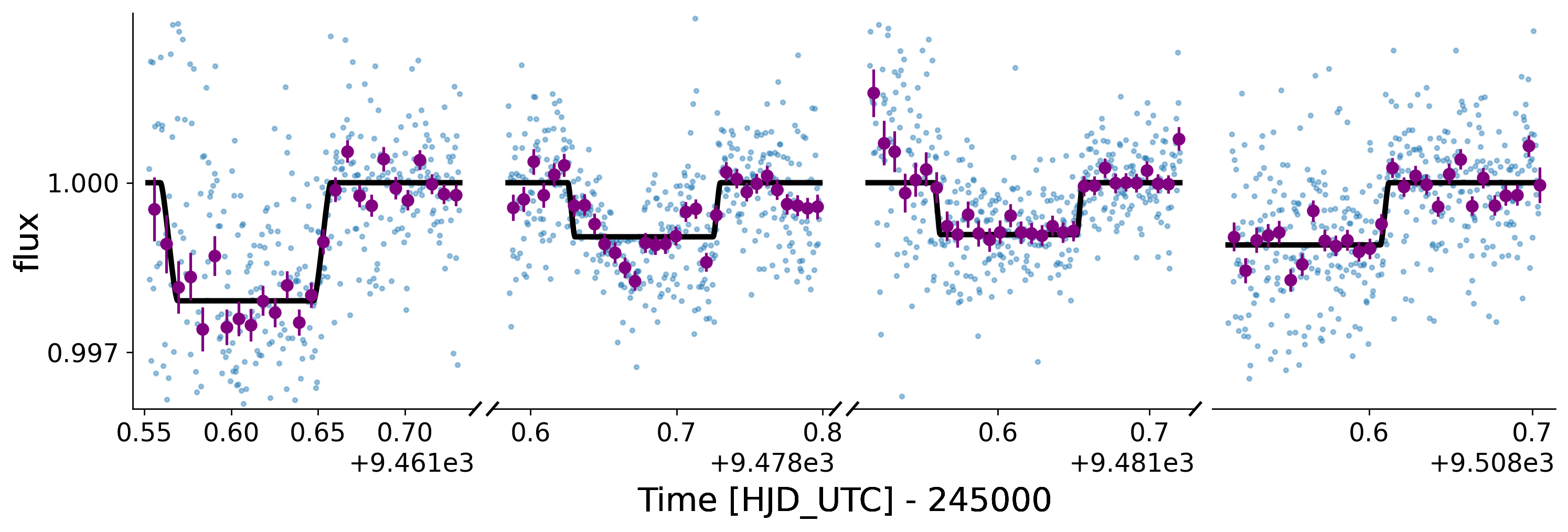}
\caption{Detrended light curves from EulerCam as described in Sec. \ref{sec:EulerCam}. Un-binned data are shown as blue points, and data in 30 min bins are shown as purple circles. The best fitting transit model for the system is shown in black.
}
\label{fig:EulerCam} 
\end{center}
\end{figure}

The detrended Eulercam data is shown in Fig. \ref{fig:EulerCam}

\section{Posteriors of stellar and noise parameters}

The posteriors of the stellar and noise parameters are given in tables \ref{tab:stellarParam_1} and \ref{tab:stellarParam_2}.

\begin{table*} 
\begin{small} 
\caption{Fitted properties of the star and noise parameters} 
\label{tab:stellarParam} 
\centering 
\begin{tabular}{llll} 
\hline 
Parameter &  Prior & photo-dynamical & photo-dynamical + RV   \\ 
\hline\hline 
\multicolumn{3}{c}{Star}\\ 
\hline 
$\rho_{\star}$ [$\rho_\odot$]&$\mathcal{N}$(2.23,0.14)&$2.244_{-0.019}^{+0.006}$&$2.34_{-0.10}^{+0.07}$ \\ 
$R_{\star}$ [$R_\odot$]&$\mathcal{N}$(0.66,0.01)&$0.6659_{-0.0075}^{+0.0039}$&$0.6592_{-0.0094}^{+0.0084}$ \\ 
\hline\hline 
\multicolumn{4}{c}{TESS}\\ 
\hline 
$\sigma$&$\mathcal{U}$(0.0e+00,1.0e+30)&$1.7e-05_{-1.1e-05}^{+1.7e-05}$&$1.7e-05_{-1.1e-05}^{+1.5e-05}$ \\ 
$u1$&$\mathcal{N}$(0.49,0.02)&$0.485_{-0.009}^{+0.011}$&$0.485_{-0.018}^{+0.015}$ \\ 
$u2$&$\mathcal{N}$(0.19,0.05)&$0.144_{-0.005}^{+0.021}$&$0.167_{-0.035}^{+0.047}$ \\ 
\hline\hline 
\multicolumn{4}{c}{CHEOPS}\\ 
\hline 
$\sigma$&$\mathcal{U}$(0.0e+00,1.0e+30)&$3.0e-04_{-5.3e-06}^{+5.9e-06}$&$3.0e-04_{-5.8e-06}^{+5.8e-06}$ \\ 
$u1$&$\mathcal{N}$(0.54,0.02)&$0.550_{-0.018}^{+0.006}$&$0.551_{-0.022}^{+0.019}$ \\ 
$u2$&$\mathcal{N}$(0.16,0.03)&$0.153_{-0.007}^{+0.017}$&$0.174_{-0.022}^{+0.024}$ \\ 
\hline\hline 
\multicolumn{4}{c}{NGTS}\\ 
\hline 
$\sigma$&$\mathcal{U}$(0.0e+00,1.0e+30)&$4.0e-05_{-2.5e-05}^{+3.1e-05}$&$3.9e-05_{-2.7e-05}^{+3.3e-05}$ \\ 
$u1$&$\mathcal{N}$(0.53,0.03)&$0.524_{-0.012}^{+0.017}$&$0.514_{-0.021}^{+0.030}$ \\ 
$u2$&$\mathcal{N}$(0.17,0.04)&$0.174_{-0.014}^{+0.008}$&$0.139_{-0.030}^{+0.037}$ \\ 
\end{tabular} 
\tablefoot{  
\label{tab:stellarParam_1} Stellar and noise parameters, with $\sigma$ the jitter terms and $uk$ the limb-darkening coefficients. }
\end{small} 
\end{table*}

\begin{table*} 
\begin{small} 
\caption{Fitted properties of the star and noise parameters} 
\label{tab:stellarParam} 
\centering 
\begin{tabular}{llll} 
\hline 
Parameter &  Prior & photo-dynamical & photo-dynamical + RV   \\ 
\hline 
offset RV&$\mathcal{U}$(-1.0e+300,1.0e+300)&$-$&$57159.10_{-0.16}^{+4.14}$ \\ 
offset $H\alpha$&$\mathcal{U}$(-1.0e+300,1.0e+300)&$-$&$0.3242_{-0.0081}^{+0.0084}$ \\ 
offset FWHM&$\mathcal{U}$(-1.0e+300,1.0e+300)&$-$&$6729.20_{-0.22}^{+0.79}$ \\ 
$\sigma$ RV&$\mathcal{U}$(0.0e+00,3.0e+01)&$-$&$0.94_{-0.11}^{+0.30}$ \\ 
$\sigma$ H$\alpha$&$\mathcal{U}$(0.0e+00,3.0e+01)&$-$&$0.00107_{-6.5e-04}^{+4.6e-04}$ \\ 
$\sigma$ FWHM&$\mathcal{U}$(0.0e+00,3.0e+01)&$-$&$3.84_{-0.16}^{+0.81}$ \\ 
$GP \alpha_{000}$&$\mathcal{N}$(0.0e+00,2.0e+02)&$-$&$10.83_{-5.49}^{+0.06}$ \\ 
$GP \alpha_{010}$&$\mathcal{N}$(0.0e+00,2.0e+02)&$-$&$0.003_{-0.021}^{+0.021}$ \\ 
$GP \alpha_{011}$&$\mathcal{N}$(0.0e+00,2.0e+02)&$-$&$-0.003_{-0.022}^{+0.020}$ \\ 
$GP \alpha_{020}$&$\mathcal{N}$(0.0e+00,2.0e+02)&$-$&$-12.06_{-4.27}^{+0.15}$ \\ 
$GP \alpha_{021}$&$\mathcal{N}$(0.0e+00,2.0e+02)&$-$&$21.57_{-0.49}^{+0.25}$ \\ 
$GP \alpha_{100}$&$\mathcal{N}$(0.0e+00,2.0e+02)&$-$&$16.35_{-1.25}^{+0.28}$ \\ 
$GP \alpha_{110}$&$\mathcal{N}$(0.0e+00,2.0e+02)&$-$&$0.016_{-0.014}^{+0.013}$ \\ 
$GP \beta_{110}$&$\mathcal{N}$(0.0e+00,2.0e+02)&$-$&$-0.014_{-0.013}^{+0.012}$ \\ 
$GP \alpha_{111}$&$\mathcal{N}$(0.0e+00,2.0e+02)&$-$&$-9.1e-04_{-0.009}^{+0.010}$ \\ 
$GP \alpha_{120}$&$\mathcal{N}$(0.0e+00,2.0e+02)&$-$&$28.73_{-0.05}^{+2.06}$ \\ 
$GP \beta_{120}$&$\mathcal{N}$(0.0e+00,2.0e+02)&$-$&$-56.46_{-0.19}^{+0.57}$ \\ 
$GP \alpha_{121}$&$\mathcal{N}$(0.0e+00,2.0e+02)&$-$&$-12.46_{-0.13}^{+1.92}$ \\ 
$GP \beta_{121}$&$\mathcal{N}$(0.0e+00,2.0e+02)&$-$&$-9.09_{-1.64}^{+0.42}$ \\ 
$GP \alpha_{200}$&$\mathcal{N}$(0.0e+00,2.0e+02)&$-$&$-2.33_{-0.97}^{+0.35}$ \\ 
$GP \alpha_{210}$&$\mathcal{N}$(0.0e+00,2.0e+02)&$-$&$-0.008_{-0.014}^{+0.009}$ \\ 
$GP \beta_{210}$&$\mathcal{N}$(0.0e+00,2.0e+02)&$-$&$0.006_{-0.011}^{+0.012}$ \\ 
$GP \alpha_{211}$&$\mathcal{N}$(0.0e+00,2.0e+02)&$-$&$-0.005_{-0.016}^{+0.013}$ \\ 
$GP \alpha_{220}$&$\mathcal{N}$(0.0e+00,2.0e+02)&$-$&$-14.33_{-0.08}^{+4.16}$ \\ 
$GP \beta_{220}$&$\mathcal{N}$(0.0e+00,2.0e+02)&$-$&$29.22_{-4.11}^{+0.12}$ \\ 
$GP \alpha_{221}$&$\mathcal{N}$(0.0e+00,2.0e+02)&$-$&$-38.19_{-0.79}^{+1.16}$ \\ 
$GP \beta_{221}$&$\mathcal{N}$(0.0e+00,2.0e+02)&$-$&$-11.72_{-0.23}^{+3.19}$ \\ 
$GP \alpha_{300}$&$\mathcal{N}$(0.0e+00,2.0e+02)&$-$&$12.43_{-2.97}^{+0.14}$ \\ 
$GP \alpha_{310}$&$\mathcal{N}$(0.0e+00,2.0e+02)&$-$&$0.009_{-0.010}^{+0.010}$ \\ 
$GP \beta_{310}$&$\mathcal{N}$(0.0e+00,2.0e+02)&$-$&$-0.008_{-0.012}^{+0.011}$ \\ 
$GP \alpha_{311}$&$\mathcal{N}$(0.0e+00,2.0e+02)&$-$&$1.3e-04_{-0.0031}^{+0.0033}$ \\ 
$GP \alpha_{320}$&$\mathcal{N}$(0.0e+00,2.0e+02)&$-$&$53.01_{-3.67}^{+0.31}$ \\ 
$GP \beta_{320}$&$\mathcal{N}$(0.0e+00,2.0e+02)&$-$&$-24.08_{-0.64}^{+0.89}$ \\ 
$GP \alpha_{321}$&$\mathcal{N}$(0.0e+00,2.0e+02)&$-$&$0.68_{-0.15}^{+4.75}$ \\ 
$GP \beta_{321}$&$\mathcal{N}$(0.0e+00,2.0e+02)&$-$&$-5.31_{-3.48}^{+0.12}$ \\ 
$GP \alpha_{400}$&$\mathcal{N}$(0.0e+00,2.0e+02)&$-$&$-6.51_{-0.85}^{+0.51}$ \\ 
$GP \alpha_{410}$&$\mathcal{N}$(0.0e+00,2.0e+02)&$-$&$-0.0084_{-0.0084}^{+0.0081}$ \\ 
$GP \beta_{410}$&$\mathcal{N}$(0.0e+00,2.0e+02)&$-$&$-0.0043_{-0.0060}^{+0.0071}$ \\ 
$GP \alpha_{411}$&$\mathcal{N}$(0.0e+00,2.0e+02)&$-$&$0.0025_{-0.0051}^{+0.0051}$ \\ 
$GP \alpha_{420}$&$\mathcal{N}$(0.0e+00,2.0e+02)&$-$&$-9.96_{-0.28}^{+0.27}$ \\ 
$GP \beta_{420}$&$\mathcal{N}$(0.0e+00,2.0e+02)&$-$&$7.21_{-0.99}^{+0.55}$ \\ 
$GP \alpha_{421}$&$\mathcal{N}$(0.0e+00,2.0e+02)&$-$&$9.25_{-1.16}^{+0.32}$ \\ 
$GP \beta_{421}$&$\mathcal{N}$(0.0e+00,2.0e+02)&$-$&$1.26_{-0.10}^{+2.55}$ \\ 
GP P&$\mathcal{LU}$(20.00,60.00)&$-$&$59.75_{-0.55}^{+0.10}$ \\ 
GP $\rho$&$\mathcal{LU}$(10.00,100000.00)&$-$&$1051.07_{-4.97}^{+0.16}$ \\ 
\end{tabular} 
\tablefoot{  
\label{tab:stellarParam_2} Noise parameters for the RV fit. $\sigma$ are the jitter term for the velocity and indicators. $P$ is the period of the GP. $\rho$ is the decay timescale. $\alpha_{klm}$ and $\beta_{klm}$ are amplitudes of the cosine and sine terms in the spleaf.term.MultiFourierKernel (\url{ https://obswww.unige.ch/\~delisle/spleaf/doc/\_autosummary/spleaf.term.MultiFourierKernel.html\#spleaf.term.MultiFourierKernel }), respectively, with $k$ the harmonic (f = k/P), l the timeseries, and m the mode.}
\end{small} 
\end{table*}

\section{Robustness of RV mass estimation and activity modeling}
\label{ap:RV}

\begin{figure*}
\begin{center}
\includegraphics[width=0.99\linewidth]{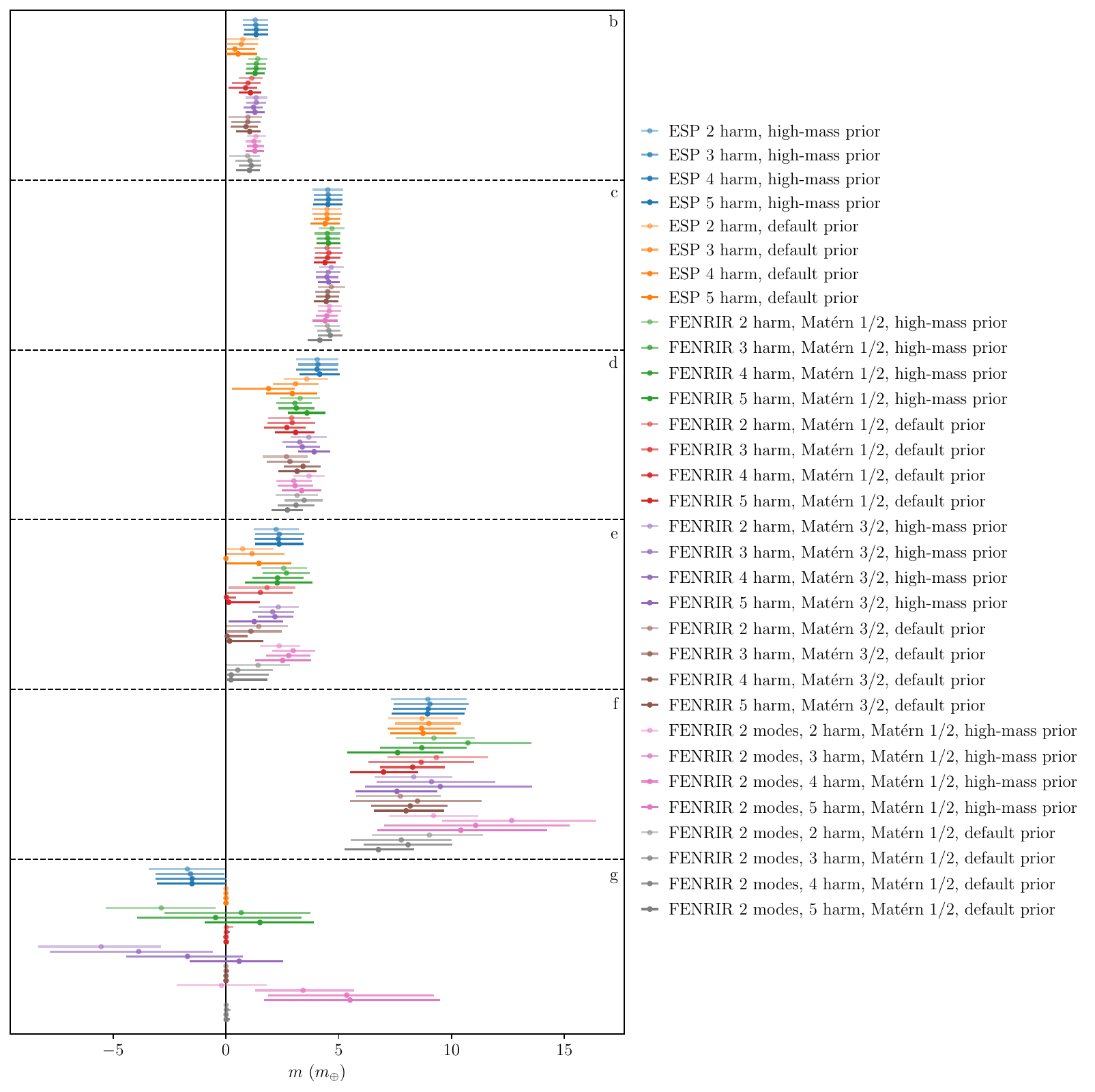}
\caption{Comparison of RV mass estimates for various activity models and priors.
}
\label{fig:TOI178_rv_noise_models} 
\end{center}
\end{figure*}

Poorly modelled noise sources can introduce biases, in particular in parameters which are always positive such as mass and eccentricity \citep{Hara2019}. 
In order to test the robustness of the mass estimates of the six planets obtained from the ESPRESSO RVs,
we analyzed the RV time series using different activity models and different mass priors.
The results are shown in Fig.~\ref{fig:TOI178_rv_noise_models}.
In all cases, we fixed the periods and phases of the planets to the values of \citet[][median values from Tables~3 and 4]{Delrez2023}.
We assumed circular orbits and only let the semi-amplitudes of the RV signals to vary.
We tested each activity model with a log-uniform (default) prior on the semi-amplitudes
and with a uniform (high-mass) prior.
In the case of the high-mass prior,
we also allow for negative semi-amplitudes (hence negative planetary masses).
A negative semi-amplitude means that the signal is found in opposite phase with what is expected from the transit.
We allow for this to better assess the reliability of the mass constraints we get from RVs and the dependency of the results with respect to the activity model and the mass prior.
We used the spleaf GP package \citep{Delisle2020}, with either the ESP kernel \citep{Delisle2022}
or the FENRIR kernel \citep{Hara2023}.
In both cases, the GP is trained simultaneously on the RVs, the FWHM, and the H$\alpha$ activity indicators.

The ESP kernel is an approximation of the widely spread SEP (squared-exponential periodic) kernel \citep[e.g.][]{Aigrain2012}.
The approximation is truncated at a given number of harmonics of the rotation period.
We used from 2 (rotation period + semi-period) to 5 harmonics.

The FENRIR kernel is, in its data-driven form that we used here, a more flexible kernel
which is also decomposed into harmonics of the rotation period, but with free coefficients.
As for the ESP kernel we used from 2 to 5 harmonics.
We also tested different decay envelopes (Matérn 1/2 and 3/2), as well as one or two modes.
Having two modes is equivalent to having two independent FENRIR processes with the same hyper-parameters but different amplitudes.
It is thus a much more flexible model than the one-mode FENRIR kernel.

From Fig.~\ref{fig:TOI178_rv_noise_models}, we conclude that the mass of the three inner planets
seems robust to changes in the activity model and in the mass prior.
The mass of planet~e is more prior dependent.
Finally planets~f and g are both prior and activity model dependent.
The mass of planet~g seems especially challenging to constraint from the existing RV data.
Some models (e.g., ESP kernel, FENRIR with a Matérn 3/2 envelope) prefer a negative mass for
planet~g, which means that they probably use the planetary signal in opposite phase
to model a component of the stellar activity.
When only allowing for positive masses, the mass posteriors for these models is still pushed toward zero.
Obtaining a negative mass is a clear diagnostic that the model is not behaving correctly,
but a bias in the opposite direction (i.e., artificially increasing the planet mass) is much more difficult to detect.
Planet~f might be in this latter case (i.e., overestimated mass).

In order to avoid biasing the final mass estimates for the combined RV + photodynamical fit,
but still benefiting from the RV constraints,
we select from this analysis the FENRIR 2 modes, 4 harm, Matérn 1/2 activity model.
Indeed, for the most model-sensitive planets (e, f, and g) this model exhibits large uncertainties on the mass posteriors and strong prior-sensitivity which reflects well the overall lack of constraint we have on these masses.
However, this does not mean that this an intrinsically better model for stellar activity.
In particular, we noticed that the rotation period is not well constrained with this model and
reaches the upper-bound of the prior interval.
We stress out here the need for additional RV measurements to better constrain the masses and activity model.

\section{Adding a planet in the chain}

\label{ap:added_planet}
\begin{figure}[!ht]
\begin{center}
\includegraphics[width=0.99\linewidth]{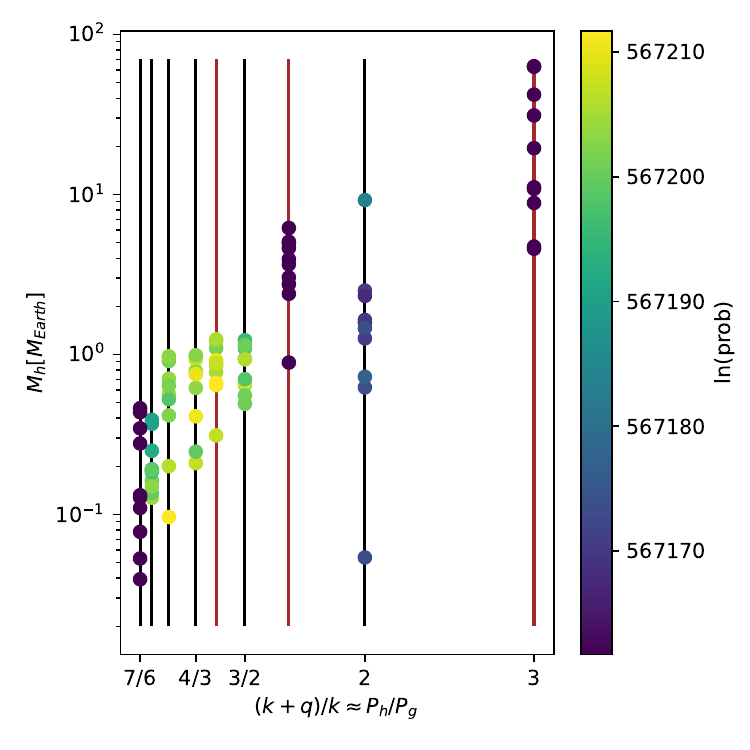}
\caption{\label{fig:TOI178h_exp} Best solutions of the joint photo-dynamical + RV fit when adding an hypothetical planet
$h$ near various MMR with respect to planet $g$.
}
\end{center}
\end{figure}

Here we check if additional planets in the resonant chain, on an external orbit to the known planet g, could either better explain the available data, or invalidate the masses that we found for the 6-planet solution. We consider a set of first- and second-order mean-motion resonances and compute what their orbital period should be to fit in the resonant chain. For each of these potential periods for planet $h$, we ran 10 fits of the individual transit timings shown in Figs. \ref{fig:TTVcde} and \ref{fig:TTVfg}. Each of these fits were initialised as follow: the orbital parameters and masses of the planets $c$ to $g$ were set to the best fit of the photo-dynamical analysis, {the} mean longitude of planet $h$ was randomly picked in the [$0^\circ$,$360^\circ$] range, the period was randomly picked in order to continue the chain with a super-period of $260\pm20$ days, see eq.C.6 of L21. $k_h=e_h \cos \varpi_h$ and $h_h=e_h \sin \varpi_h$ were initially set at $3e-4$ and $M_h/M_\sun$ was initially set to $6*10^{-7}$. The set of priors is identical to the \textit{highmass} case described in section \ref{sec:approach}. Once these fits converged, the best solution of each fit is used to initialise a joint photo-dynamical+RV fit, identical in all point to those presented in section \ref{sec:approach}, only adding the additional planet. 

In Fig. \ref{fig:TOI178h_exp}, we show the best solution of each of these fits, as a function of the MMR near which the $g$-$h$ pair is, the mass of planet $h$, and the logprob of the solution. From these, we find that several resonances have similar probabilities, with the general trend that the larger $P_h$, the more massive $m_h$ is. Such degeneracy is well understood when considering the TTV induced at the super-period for nearly-resonant pair of planets \citep[e.g.][]{Lithwick2012,HaLi2016}. However, there are additional arguments that could have allowed us to favour some solutions with respect to others. Indeed, small period ratios can also lead to short-term sawtooth-like TTVs known as chopping \citep[e.g.]{Deck2015}, that could better explain the TTVs of planet $g$. In addition, some of the tried periods for $h$ are also close to significant MMR with planet $f$: for example, $g$ and $h$ close to a $3:2$ or a $5:4$,  MMR implies that $f$ and $h$ are close to a $2:1$ or a $5:3$ MMR, which could also impact the overall goodness of the fit. 

Our best solutions are for an additional planet between a 5/4 and 3/2 MMR with planet $g$, and masses of the order of 1$M_{Earth}$ or less. We would like to point out that the parameter space for this potential additional planet is much larger than the ones explored here, and therefore we cannot exclude that an additional more massive planet is orbiting TOI-178, especially if it is not in direct continuation of the resonant chain. Nonetheless, this preliminary study allows us to draw two conclusions: there are no obvious signals in the available data for an additional planet in the chain, and the best fits of the available data with the additional planet do not significantly impact the derived mass for the known planets in the system. Indeed, we aggregated the mass of the planets $b$ to $g$ for the best fits of the configurations in which the added planet has a period in either 5/4, 4/3, 7/5 or 3/2 MMR with planet $g$. These aggregated results are represented in cyan in the Fig. \ref{fig:masspost}. As one can see, these aggregated masses remain close to the 6-planet solution with the \textit{highmass} prior.


 

\end{document}